\documentclass[12pt,preprint]{aastex}
 
\newcommand{\rrab}{RR{\it ab} }
\newcommand{\rrc}{RR{\it c} }

\slugcomment{To appear in the AJ}
 
\shorttitle{QUEST RR Lyrae Survey}
\shortauthors{Vivas \& Zinn}
 
\begin{document}
 
\title{The QUEST RR Lyrae Survey II: The Halo Overdensities in the First Catalog}

\author{A. Katherina Vivas}
\affil{Centro de Investigaciones de Astronom{\'\i}a (CIDA). Apartado 
Postal 264. M\'erida 5101-A. Venezuela}
\email{akvivas@cida.ve}

\and
\author{Robert Zinn}
\affil{Yale University. Astronomy Department. PO Box 208101. New Haven, 
	CT 06511}
\email{zinn@astro.yale.edu}

\begin{abstract}
The first catalog of the RR Lyrae stars (RRLS) in the Galactic halo by the 
Quasar Equatorial Survey Team (QUEST) has been searched for significant 
overdensities that may be debris from disrupted dwarf galaxies or 
globular clusters.  These RRLS are contained in a band $\sim 2\fdg 3$  
wide in declination that spans $\sim 165\degr$ in right ascension and lie 
$\sim 4$ to $\sim 60$ kpc from the Sun.  Away from the major overdensities, 
the distribution of these stars is adequately fit by a smooth halo model, 
in which the flattening of the halo decreases with increasing 
galactocentric distance \citep{pre91}.  
This model was used to estimate the ``background'' 
of RRLS on which the halo overdensities are overlaid.  A procedure was developed 
for recognizing groups of stars that constitute significant overdensities 
with respect to this background.  To test this procedure, a Monte Carlo 
routine was used to make artificial RRLS surveys that follow the smooth halo 
model, but with Poisson distributed noise in the numbers of RRLS and, within 
limits, random variations in the positions and magnitudes of the artificial 
stars.  The $10^4$ artificial surveys created by this routine were examined 
for significant groups in exactly the same way as the QUEST survey.  These 
calculations provided estimates of the frequencies with which random fluctuations 
produce significant groups.

In the QUEST survey, there are six significant overdensities that contain six or 
more stars and several smaller ones.  The small ones and possibly one or two of 
the larger ones may be artifacts of statistical fluctuations, and they need to be 
confirmed by measurements of radial velocity and/or proper motion.  The most 
prominent groups are the northern stream from the Sagittarius dwarf spheroidal 
galaxy and a large group in Virgo, formerly known as the ``12.4 hr clump'', 
which recently \citet{duf06} have shown contains a stellar stream 
(the Virgo Stellar Stream).  Two other groups lie in the direction of the Monoceros 
stream and at approximately the right distance for membership.  
Another group is related to the globular cluster Palomar 5.

\end{abstract}
 
 
\section{INTRODUCTION}

Recent surveys of the galactic halo have shown that it contains
streams of stars emanating from dwarf galaxies.  The firmest evidence
for this comes from the numerous detections of streams from the 
Sagittarius (Sgr)
dwarf spheroidal galaxy (hereafter dSph), which wrap around the sky
\citep{maj03}.  
There is also strong evidence for another large stream, the Monoceros Stream 
\citep{new02,yan03}, that forms a ring-like structure around the Milky Way 
\citep{iba03}. Several authors have argued that the parent galaxy of the Mon 
Stream lies in the constellation Canis Major, but there is considerable controversy 
over this association and even the existence of the CMa galaxy 
\citep[among others]{mar04,mom04,m-d05,car05,pen05}.
These on-going merger events may be only the most recent examples of a long history
of mergers that build the Milky Way from an ensemble of smaller
systems, as proposed by hierarchical picture of galaxy formation \citep[e.g.][]{bul05}.
While the evidence for ancient merger events are less spectacular than
that for the recent ones, it is nonetheless compelling.  It has long
been speculated that the thick disk was produced by a merger of the
Milky Way with a relatively large satellite galaxy soon after the
formation of the first disk structure \citep[see review by][]{fre02}.  
Although the very unusual
globular cluster $\omega$Cen has not yet been directly linked to a
merger event, its exceptional structure and internal ranges in
metallicity and in age has fueled speculation that it is the nucleus
of a now extinct nucleated dwarf galaxy 
\citep[][among others]{car00,tsu03,ide04,rey04}.

Simulations of the destruction of dwarf galaxies \citep{joh96,har01} 
have shown that
the tidal debris may be recognized in the halo as over-densities in
space and as coherent structures in radial velocity space long after
the merger event. By detecting this debris, modeling their orbits, and
studying their ages and chemical compositions, the merger history of
the Milky Way may be pieced together.  In addition to documenting the
importance of mergers to galactic evolution, this may help explain the
large inconsistency between the number of satellite galaxies predicted
by the hierarchical picture for the formation of a large galaxy and
the relatively small number of satellite galaxies around the Milky
Way, the "satellite problem" \citep[see][]{fre02}.

Debris from satellite galaxies is not the only kind expected in the
galactic halo.  Numerical simulations have shown that several
processes can lead to the disruption of globular clusters, and there
has been much speculation that the roughly 150 the globular clusters
that are now identified in the Galaxy are the survivors of a once much
larger population \citep[e.g.][]{gne97}.  A few examples of tidal tails from globular
clusters have been reported \citep{gri95,leo00}, the most spectacular of which are the
streams that stretch several degrees from the globular cluster 
Pal 5 \citep{ode01,ode03}.
\citet{kin04} have detected some small groups of RR Lyrae
variables that have similar radial velocities and metallicities,
precisely the properties expected of debris from disrupted globular
clusters.

In this paper we discuss the spatial distribution of the stars that
were discovered in the first band of the QUEST RR Lyrae Survey \citep{viv04}.
RR Lyrae stars (hereafter RRLS) were selected for this
survey of halo substructure because they are easily detected and are
excellent standard candles.  Our primary goal is to identify
overdensities that may be debris from dwarf galaxies or from globular
clusters.  In a few structures identified here, the densities of the
RRLS are so much higher than the average densities of variables
that there is little doubt that the feature is real and not a random
fluctuation in the background.  Not all of the other features
identified here may be real, and they require confirmation by
determining if their member stars have similar motions.  We will
report in later papers our measurements of the radial velocities and
the metallicities of the stars in some of these spatial groups.

\section{THE POPULATION BIAS OF A RRLS SURVEY}
\label{sec:frequency}

Before discussing the QUEST survey, it is import to consider a bias that 
affects all RRLS surveys. RRLS are only found in the oldest 
(age $\geq 9$ Gyrs) stellar populations that have the proper combination 
of [Fe/H] and other factors (``2nd parameters'' e.g.,  age, CNO/Fe, He/H, 
and stellar density) to enable horizontal branch (HB) stars to evolve into 
the instability strip.  These complexities of HB morphology produce 
a wide range of RRLS populations even among the outwardly similar old and 
metal-poor globular clusters in the Galactic halo. Some of these clusters 
contain several tens of RRLS, others contain only a few RRLS, and some 
contain none at all. Clearly a search for the debris from destroyed globular 
clusters is limited by this effect, but does it also seriously impact a 
search for stellar streams from dwarf galaxies?

The RRLS populations in globular clusters has been quantified by \citet{sun91}, who
following \citet{kuk73}, calculated the specific frequency of RRLS
which they defined as the number of RRLS per unit absolute visual magnitude ($M_V$),
normalized to $M_V = -7.5$ (we call this $S_{RR}$ following the 2003 version of the 
\citet{har96} catalogue). This fiducial $M_V$ is near the peak of the distribution 
of the Milky Way globular clusters. Using measurements of $S_{RR}$ for a 
large sample of globular clusters, \citet{sun91} demonstrated that there is a systematic
variation in the frequencies of RRLS with Galactocentric distance ($R_{gal}$) in 
the sense that the metal-poor globular clusters in the outer halo 
($R_{gal} > R_0$) have higher frequencies than do inner halo clusters of the same
metallicity. This effect, which is due to the variation of the 2nd parameter with
$R_{gal}$ \citep[e.g.][]{sz78} may explain the difference in mean metallicity 
between the field RRLS lying in the inner and outer halos \citep{sun91,zin86}.

We are interested here in examining $S_{RR}$ in low-mass dwarf galaxies as well as in
globular clusters.  The dSph galaxies provide a convenient sample of dwarf galaxies 
covering the extreme low-mass range of the distribution of galaxies. Because they are the
most numerous type of satellite galaxy around the Milky Way and M31, it is likely that 
dSph galaxies  were the type most frequently destroyed in the past.  The Sgr dSph 
galaxy is of course undergoing tidal destruction at the present time. Values of $S_{RR}$ 
for 9 of the 10 known dSph companions of the Milky Way and for 4 of the 7 dSph 
companions of M31 have been computed and are listed in Table~\ref{tab-frequency}
(insufficient data were available for the other dSph companions).  With two exceptions, 
the values of $M_V$ in this table were taken from \citet{irw95} and \citet{mcc06} 
for the Milky Way and the And systems, respectively. \citet{ode01} have shown that 
Draco has a significantly larger radius than was previously measured.  To estimate  
the $M_V$ of Draco, we adopted $(V-I_C)_0 = 0.9$ for its integrated color, which is
consistent with its old and very metal-poor stellar population.  With the 
transformation equation in \citet{ode01}, this value yields $(V-i)_0 = 0.4$, 
which we added to the value of $M_i = -8.77 \pm 0.2$ \citep{ode01}. For the Sgr dSph, 
we adopted the $M_V$ that \citet{cse01} derived, which lies in near the middle of the 
range quoted in the recent literature. The values of the mean [Fe/H] of the galaxies 
were taken from \citet{mat98} for the Milky Way companions and from 
\citet{pri02,pri04,pri05} for the And galaxies. Because dSph galaxies have 
significant internal dispersions in [Fe/H] and because the RRLS are among 
the oldest stars, the mean [Fe/H] of a galaxy may be larger than the mean 
value of its RRLS. The important point for our discussion is that the RRLS in these 
systems probably span a range in [Fe/H] that is not grossly different from the range
observed among the RRLS in the outer Galactic halo 
\citep[$-0.9 > \rm{[Fe/H]} > -2.5$]{sun91}. It is possible then that the field RRLS 
formed in similar galaxies that were later torn apart. In Table~\ref{tab-frequency}, 
$N_{obs}$ is the number of RRLS that were observed in the fields covered in the 
variability searches, and $N_{est}$ is our estimate of the total number of RRLS in the
galaxies. For most of the galaxies, we calculated $N_{est}$ from $N_{obs}$ 
by dividing $N_{obs}$ by the fraction of total galaxy light that is included in the 
observed field. For the Milky Way systems, we computed these fractions from \citet{kin62}
profiles, using the parameters given by \citet{irw95}.
The ellipticities of the galaxies were taken into account, and it was assumed that the
observed fields coincided with the centers of the systems.  \citet{pri02,pri04,pri05}
estimated the fractions of the luminosities of the And galaxies that were covered by their
fields, which made the calculations of $N_{est}$ straightforward. For the Sgr dSph galaxy, 
we adopted the estimate by \citet{cse01} that its main body contains 4200 type ab variables.
We then used the observed ratio of the type c to type ab RRLS to derive $N_{est}$.  
For Draco, we set $N_{est} = N_{obs}$ because \citet{kin02} did not specify the size of
their field. Draco is so rich in variables that it matters little to our discussion that
this procedure may have underestimated its $S_{RR}$.  The standard deviations ($\sigma$) 
of the values of $S_{RR}$ were calculated by assuming Poisson statistics applies to 
$N_{est}$ and by adopting the uncertainties in $M_V$ given by their sources.
Because the $N_{est}$ values are only estimates, the $\sigma$'s listed in
Table~\ref{tab-frequency} should be treated as lower limits. The variability studies
probably missed some variables, and  consequently, the values of $S_{RR}$ in 
Table~\ref{tab-frequency} are more likely to be too small than too large. 
This is particularly true for the Fornax galaxy because \citet{ber02} state there are other
variable stars of the right magnitude to be RRLS for which they could not determine periods.

\begin{figure}
\epsscale{0.83}
\plotone{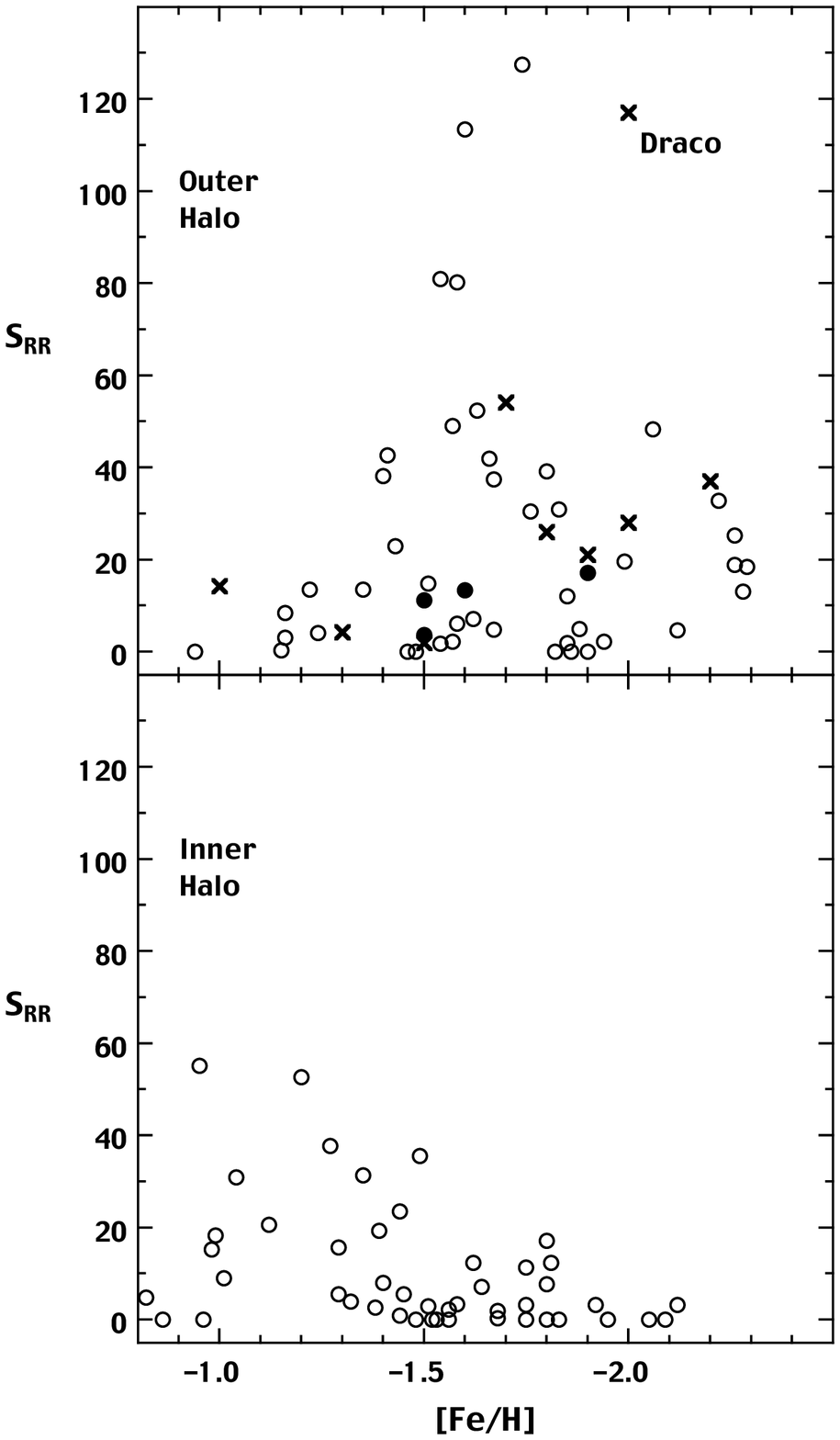}
\caption{For halo globular clusters (open circles) and dSph companions of
the Milky Way ($\times$'s) and Andromeda (solid points), the specfic frequency of 
RRLS ($S_{RR}$) is plotted against [Fe/H].  The upper and lower diagrams are 
the plots for the outer ($R_{gal} > R_0$) and inner ($R_{gal} < R_0$) halo, 
respectively.  All of the Milky Way dSph companions lie in the outer halo, 
where we have also plotted the And galaxies for comparison.}
\label{fig-nrr}
\end{figure}

In Figure~\ref{fig-nrr}, $S_{RR}$ for the globular clusters (Harris 2003 catalogue)
and the dSph galaxies are compared. One can see again the difference in $S_{RR}$ 
among the metal-poor globular clusters in the inner and outer halos. The near zero values 
of $S_{RR}$ among the metal-poor clusters of the inner halo suggest that RRLS may not be a
good tracer of its stellar populations. RRLS may be better probes of the outer halo, where
the mean value of $S_{RR}$ for the clusters is well above zero. The dSph galaxies span a
wide range in $S_{RR}$, but none has $S_{RR} = 0$. Draco has a remarkably high value that 
is comparable to the highest values in the sample of 90 globular clusters.
Several of the other dSph galaxies have values of 
$S_{RR}$ that are comparable to the variable-rich outer halo clusters.
This is somewhat remarkable because many of these galaxies contain substantial populations
of intermediate-aged stars that contribute to the total luminosities of the galaxies but not
of course to the samples of RRLS.

\begin{figure}
\epsscale{0.88} 
\plotone{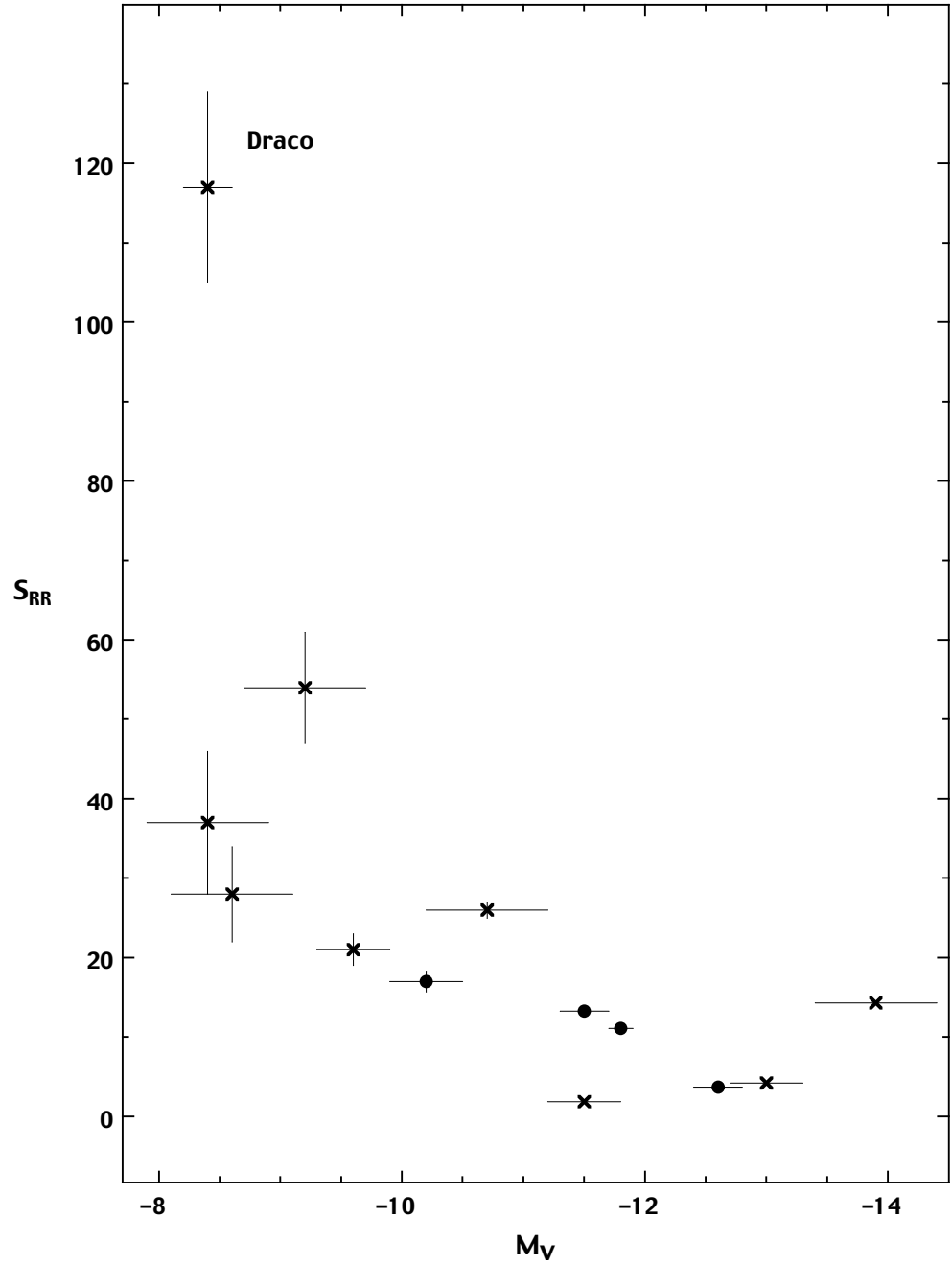}
 \caption{The specific frequencies ($S_{RR}$) of the RRLS in dSph galaxies is 
plotted against their absolute visual magnitudes ($M_V$).  
The $\times$'s and solid points depict the Milky Way and Andromeda companions, 
respectively.}
 \label{fig-nrrMv}
\end{figure}

$S_{RR}$ is plotted against $M_V$ for the galaxies in Figure~\ref{fig-nrrMv}, 
where one can see a clear trend of decreasing $S_{RR}$ with increasing galaxy luminosity.
This is not surprising because the more luminous galaxies tend to have larger populations 
of intermediate-age stars. Note that the Milky Way and And galaxies are indistinguishable 
in this plot. The small but still significantly above zero values of $S_{RR}$ 
among the high luminosity galaxies indicate that their destruction will release large
numbers of RRLS. It is not surprising then that the tidal streams from the Sgr dSph galaxy
have been easily detected by RRLS surveys \citep[see below and also][]{ive00}.
The higher $S_{RR}$ among the low luminosity dSph galaxies partially offsets the greater
difficulty that any probe of the halo will have detecting the smaller streams that are
produced by the tidal destruction of low-luminosity systems.

If the 13 galaxies in Table~\ref{tab-frequency} are representative of the dwarf galaxies
that merged with the Milky Way, then RRLS surveys for stellar streams are probably not
seriously handicapped by the population bias.  They may also detect the debris from the
destruction of the most luminous and variable rich globular clusters.  Because RRLS are
superior standard candles (see below),  RRLS surveys may provide a better description of
halo substructure than other surveys that employ probes that suffer less from population biases (e.g., K giants).

\section{THE QUEST RR LYRAE SURVEY}

The first band of the QUEST RR Lyrae survey \citep[][hereafter Paper I]{viv04}
identified 498 RRLS in almost 380 deg$^2$ of the sky that
spans a wide range of galactic coordinates and in apparent magnitude
($13.7<V<19.7$).  Paper I describes in detail the techniques of the
survey and its completeness.  The survey covered a $2\fdg 3$ wide strip
of the sky, centered at declination $\delta = -1^\circ10\farcm 8$, from right
ascencion ($\alpha$) $4\fh 1$ to $6\fh 1$ and from $8\fh 0$ to $17\fh 0$.  The span
from $6\fh 1$ to $8\fh 0$ was not observed because it includes regions near
the galactic plane. Subsequent work on the individual stars in the catalogue
has shown that 41 of them are not real variables. Their apparent variability
was produced in most cases 
by close neighbors not resolved by the QUEST instrumentation. The 
photometric pipeline that was used did not include a deblending algorithm.
The regions most affected by blends were the ones closest to the
galactic plane which present the most crowding. These regions 
were also the less observed in the QUEST survey and consequently there is
a small number of observations per star. 
With few epochs available there was a greater chance that the variations 
in magnitude that were produced by the poor centering on a blended image 
mimicked a RRLS light curve.
Only 3 cases of blends
were found at high galactic latitudes ($9\fh 1 < \alpha < 16\fh 0$),
which represents only the $0.8\%$ of all the stars in the catalogue in 
that region.We eliminated all cases of blends before preforming the following 
analysis on the remaining 457 RRLS. 

\section{THE DISTANCES OF THE RRLS}

The distances of the RRLS from the Sun ($r_\sun$) are estimated in the usual way
from measurements of their mean V magnitudes, $\langle V \rangle$, and interstellar
extinctions, $A_V$, and by assuming an average value for their
absolute magnitudes, $M_V$:

\begin{equation}
r_\sun {\rm (kpc)} = 10^{(\langle V \rangle - M_V - A_V +5)/5} \times 10^{-3}
\end{equation}
 
Because our ability to detect small halo
substructures depends on the precisions of the distances obtained, it
is essential to estimate the errors of these measurements and to
investigate the variation in $M_V$ among the RRLS.

The uncertainty in the magnitude system of the QUEST observations
(0.02 mag, see Paper I) is sufficiently small that random errors
dominate the error budget for most of the RRLS.  The values of $\langle V \rangle$ for
the type ab RRLS (\rrab) were determined by fitting template light curves from
\citet{lay98} to the QUEST observations and then integrating these curves
after they were transformed to intensity units (see Paper I).  For the
type c variables (\rrc), the mean value of the individual V observations was
adopted for $\langle V \rangle$.  The errors in the values of $\langle V \rangle$ obtained by
these methods vary from 0.01 to 0.12 and average 0.05.  This variation
is caused by a combination in the errors in the V measurements, which
increase with increasing V, and the uneven coverage in phase of the
observations.  

\begin{figure}
\plotone{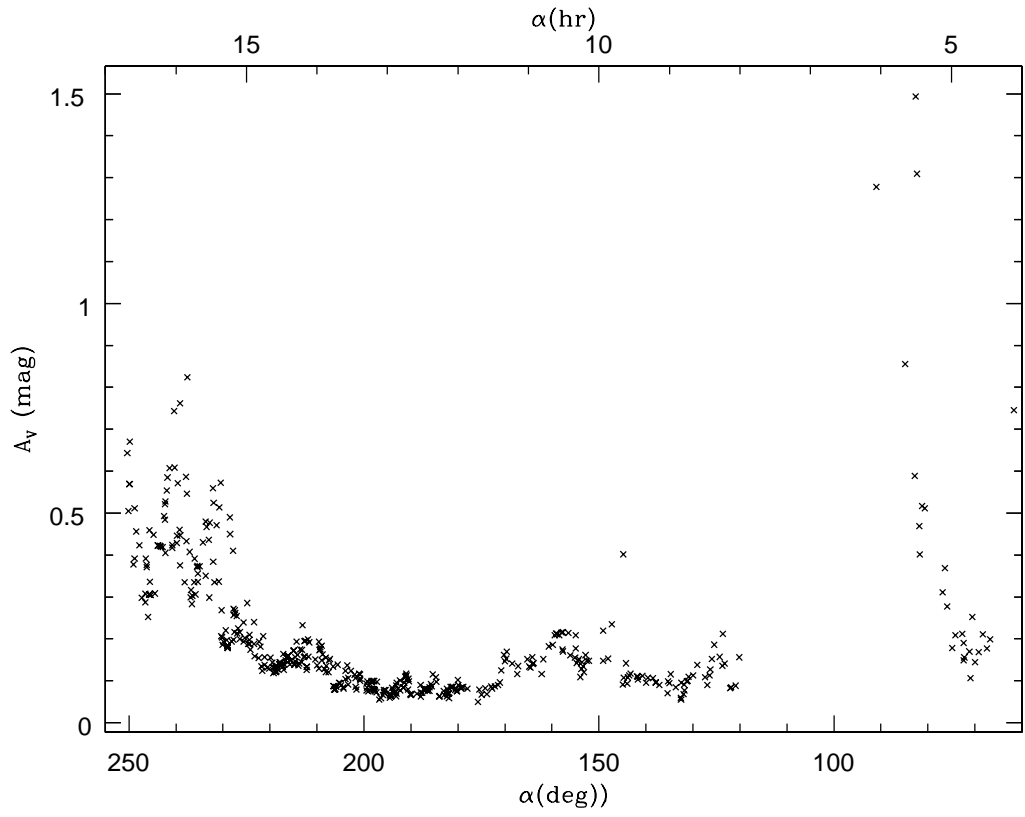}
\caption{Extinction in V magnitude of the RRLS in the QUEST survey 
plotted as a function of right ascension.}
\label{fig-ext}
\end{figure}

The interstellar reddenings of each RRLS was obtained from the dust
maps of \citet{sch98} and transformed to extinction using
$A_V=3.24 \, E(B-V)$.  Figure~\ref{fig-ext} shows the value of $A_V$ for
each star as a function of right ascension. The instellar extinction 
is low except in the directions approaching the galactic plane.  For
example, in the region between $\alpha=5-6$ h, which includes portions
of the dense molecular clouds of the Orion star forming region, some
stars have up to $\sim 1.5$ magnitudes of extinction.  Since it is very
difficult to estimate the completeness of the survey in this part of
the sky where the extinction is highly variable, this region is
excluded when we consider the shape and density fall-off of the halo.
It is examined, however, for density enhancements.  Most of our survey
is at much higher galactic latitudes where $A_V\leqslant 0.6$.  Since
\citet{sch98} estimate that the errors in their values of E(B-V) are
$\sim 10\%$, the standard deviation of $A_V$, $\sigma_{A_V}$, is probably
$\leqslant0.06$ for most stars.

The $M_V$ of a RRLS depends on its position on the evolutionary track
from the zero age horizontal branch (ZAHB) and on its metallicity,
which alters the evolutionary tracks and timescales and changes the
$M_V$ of the ZAHB.  It is convenient to summarize this as: 
$M_V = M_V^{\langle {\rm HB} \rangle} + \Delta M_V^{\rm ev}$, 
where $M_V^{\langle {\rm HB} \rangle}$ is the mean
absolute visual magnitude of the horizontal branch at the instability
strip and $\Delta M_V^{\rm ev}$ is a star's deviation from this mean value
due to the state of its evolution.  
The dispersion in these quantities are estimated in the following analysis.

\begin{figure}
\epsscale{0.79}
\plotone{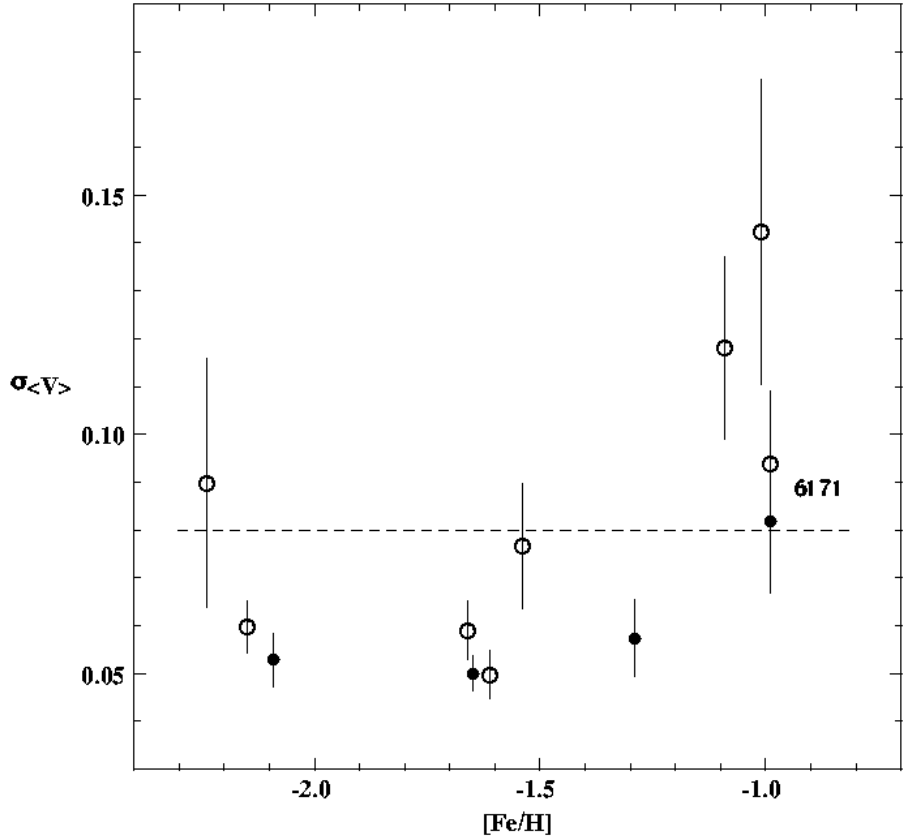}
\caption{For a sample of globular clusters, the standard deviations of the $\langle V \rangle$ 
magnitudes of their RRLS are plotted as a function of the [Fe/H] values of the clusters.  
The open symbols are the values computed from the photographic 
photometry compiled by \citet{san90}.  
The solid symbols are the values computed from the CCD photometry (see text).  Note the 
agreement between the photographic and the CCD values for NGC 6171.  The dashed horizontal 
line is the value of $\sigma_{\langle V \rangle}$ that was adopted for the dispersion 
in $\Delta M_V^{\rm ev}$.}
\label{fig-sigmaVHB}
\end{figure}

The star to star variation in $\langle V \rangle$ among the RRLS in a globular
cluster, $\sigma_{\langle V \rangle}$, provides a measure of the dispersion in
$\Delta M_V^{\rm ev}$ at fixed [Fe/H] because there is very little
variation in [Fe/H] among the stars in a typical globular cluster.  It
is essential to calculate $\sigma_{\langle V \rangle}$ for a number of clusters that
span a range in [Fe/H] because of the changes in track morphology
makes $\sigma_{\langle V \rangle}$ larger for the metal rich clusters \citep[see][]{san90}.
Figure~\ref{fig-sigmaVHB} shows the dependence of $\sigma_{\langle V \rangle}$ on [Fe/H],
where we have plotted the values of $\sigma_{\langle V \rangle}$ that we calculated
for the clusters M3, M15, M92, NGC 3201, 6171, 6712, 6723 and 6981
from the photographic photometry compiled by \citet{san90}.  We
ignored Sandage's data for $\omega$Cen because its RRLS vary in [Fe/H]
and for M4 because it has variable interstellar extinction across its
face \citep[][and references therein]{liu90,iva00}, which
significantly increases the value of $\sigma_{\langle V \rangle}$.  Accurate
background subtraction is notoriously difficult in the crowded
fields of globular clusters.  Since it is done differently by the
techniques of photographic and CCD photometry, we measured a few
values of $\sigma_{\langle V \rangle}$ for clusters that had been observed by CCD
cameras.  The CCD photometry of the RRLS in NGC 6171 by \citet{cle97}
yields a value of $\sigma_{\langle V \rangle}$ that agrees to within
the errors with the photographic value, and the values that we
obtained from the CCD photometry of M5 \citep{bro96}, IC 4499
\citep{wal96}, and M68 \citep{wal94} are consistent with the
measurements from the photographic photometry of other clusters that
have similar metallicities. In all cases, the variation in $\langle V \rangle$ among
the RRLS in a cluster is approximately Gaussian. Because very few halo
RRLS are as metal rich as [Fe/H]$\sim -1$ \citep{sun91}, we
adopt $\sigma_{\langle V \rangle} = 0.08$ for the dispersion in $\Delta M_V^{\rm ev}$.
Figure~\ref{fig-sigmaVHB} shows that this value is clearly an overestimate for all but
the most metal rich RRLS.

While it is well known that the $M_V^{\langle {\rm HB} \rangle}$ 
varies with [Fe/H], there
is much debate among different authors over the functional form of
this variation and the values of $M_V$ at particular values of [Fe/H].
This has been reviewed by \citet{cac03b}, who obtained
from a weighted average of ten methods in the literature the value of
$0.59\pm 0.03$ for $M_V^{\langle {\rm HB} \rangle}$ at [Fe/H]$=-1.5$.  
While there is
considerable evidence that $M_V^{\langle {\rm HB} \rangle}$ is a nonlinear function of
[Fe/H], over the [Fe/H] range of halo RRLS
($-2.3\lesssim {\rm [Fe/H]} \lesssim -1$), most of the recently proposed
$M_V^{\langle {\rm HB} \rangle} -$ [Fe/H] relationships can be closely approximated by a
linear dependence with a slope of $0.23\pm 0.04$ \citep{cha99,cac03a}. We have
therefore adopted the relationship:

\begin{equation}
M_V^{\langle {\rm HB} \rangle} = (0.23\pm 0.04)\, ({\rm [Fe/H]}+1.5)+(0.59\pm 0.03)
\end{equation}

Using the above $M_V^{\langle {\rm HB} \rangle} -$ [Fe/H] relationship, 
we can investigate
the distance errors introduced by the adoption of one value of $M_V$
for all RRLS.  An estimate of the likely distribution of [Fe/H] among
the QUEST RRLS is provided the spectroscopic observations by \citet{sun91}
of 113 halo RRLS lying at galactocentric distances
greater than 8.5 kpc.  The distribution of these measurements is
approximately Gaussian and has a mean value of [Fe/H]$=-1.65$ and a
standard deviation, corrected for measuring errors, of 0.30 dex.  These
values are on the \citet{zin84} metallicity scale for globular
clusters, which despite its age agrees well with recent determinations
based on high dispersion spectroscopy \citep{kra03}.  This metallicity
dispersion and the uncertainties in the $M_V^{\langle {\rm HB} \rangle}$ 
relation produce
$\sigma_{M_V^{\langle {\rm HB} \rangle}} = 0.10$.  
When combined with our estimate of 0.08
for the dispersion in $\Delta M_V^{\rm ev}$, we obtain
$\sigma_{M_V}=0.13$. Using the above estimates of the errors in the mean
magnitudes and interstellar extinctions, we obtain 0.15 for the $1\sigma$
uncertainty in the true distance modulus for a typical RRLS in
the QUEST survey.  This translates into a fractional error
($\sigma_{r_\sun}/\! r_\sun $) of only 0.07.  
The distances to most other halo tracers
have larger fractional errors, which for K giants \citep{doh01}
M giants \citep{maj03}, and
main-sequence stars \citep{jur05,new02} are
larger by factors of $\sim 2-3$.  Only blue horizontal branch (BHB)
stars, with fractional error estimates from $6-10\%$ \citep{bro05,sir04},
are similar to the RRLS in the precision of their distance estimates.  
But BHB stars must be
observed spectroscopically before they can be reliably separated
from other types of blue stars.  For determining the distances of the
QUEST RRLS, we have adopted, as we have previously \citep[e.g.][]{viv01},
$M_V=0.55$, which is consistent to 0.01 mag. with the above
$M_V-$[Fe/H] relation and the mean halo metallicity measured by \citet{sun91}.

Figure~\ref{fig-sky} shows a polar plot of the right ascension and
extinction corrected mean magnitudes $V_0$, of the RRLS in the survey.
The dotted circles indicate the
values of $V_0$ corresponding to $r_\sun= $8, 19 and 49 kpc. This simple plot
shows that the distribution of RRLS far from the Sun is not
uniform. Particularly notable is the group of stars at $V_0 \sim 19$,
$\alpha=13\fh 0 - 15\fh 4$, which was reported earlier in \citet{viv01}.
This feature, which is undoubtedly tidal debris from the Sgr dSph
galaxy, is described in more detail later along with other density
enhancements.

The positions of the RRLS in a Galactic Cartesian system have been
calculated from their galactic coordinates, $l$ and $b$, and their
heliocentric distances of the RRLS using the equations:

\begin{eqnarray}
x & = & R_0 - r_\sun \, \cos b \, \cos l \nonumber \\
y & = & r_\sun \, \cos b \, \sin l \\
z & = & r_\sun \, \sin b \nonumber
\label{eq-transf}
\end{eqnarray}

In this coordinate system, the $xy$-plane coincides with the galactic
plane, with the line from the Galactic Center (GC) to the Sun defining
the the $x$-axis.  Positive $x$ values are on the Sun's side of the GC and,
the positive $y$ axis is in the direction of $l=90\degr$.  
Coordinate $z$ is positive
towards the North Galactic Pole.  A value of 8 kpc was adopted for
$R_0$, the distance of the Sun from the GC \citep{rei93}.
Galactocentric distances are given by

\begin{equation}
R_{gal} = \sqrt{x^2 + y^2 + z^2}
\label{eq-rgal}
\end{equation}

Table~\ref{tab-rrgc} contains the galactic coordinates ($l$, $b$), extinction
corrected V magnitude ($V_0$), galactocentric
coordinates ($x$, $y$, $z$) and heliocentric and galactocentric
distances, $r_\sun$ and $R_{gal}$. The ID numbers of the stars are
the same as in Table 2 of Paper I. The full version of Table~\ref{tab-rrgc} 
is available only on-line as a machine-readable table.

\section{THE SPACE DENSITIES}

\begin{figure}
\plotone{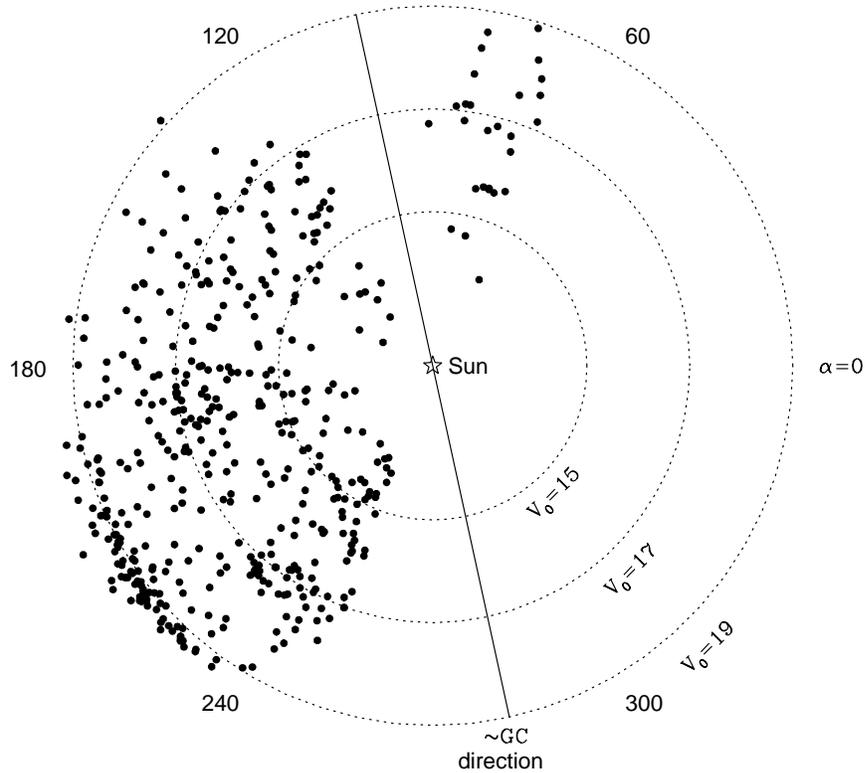}
\caption{Radial plot of the distribution of the RR Lyrae stars in right
ascension. The radial axis is the extinction corrected $\langle V \rangle$ 
magnitudes. 
All stars are located in a $2.3^\circ$ wide strip centered at declination
$-1^\circ10'.8$. The
circles correspond to a distance from the Sun of 8, 19 and 49 kpc. 
The solid line indicates the position of the galactic plane, which is located 
approximately in the GC and Galactic anticenter direction ($l \sim 32\degr$ and
$l \sim 214\degr$).
The globular cluter Pal 5 is located at $\alpha \sim 229 \degr$ and 
$\langle V_0 \rangle \sim 17.3$.}
\label{fig-sky}
\end{figure}

Although Figure~\ref{fig-sky} and several other recent halo surveys (see \S 1) have
shown that the halo does not have smooth density contours, it is still
useful to use this approximation to estimate the ``background'' of
halo RRLS upon which density enhancements, such as the one described
above, are overlaid.  In this spirit, we have performed the following
analysis. 

Any determination of the space density of RRLS requires a careful
consideration of the survey's completeness.  In Paper I we showed that
the completeness of the \rrc variables is significantly lower than that of the
\rrab, especially at the faint end of the survey. The sample of
\rrc may have also a small, but unknown, contamination from
eclipsing binaries and variable blue stragglers (see Paper I). 
For these reasons, we have used only the
\rrab to estimate the density distribution.  As we
explained above, we do not use stars in the region with $\alpha = 5 -
6$ hrs because of the large interstellar extinction. Finally, since
the saturation limit of the different CCDs in the QUEST camera is not
uniform (see Paper I), we eliminated stars with $\langle V \rangle
<15.0$. This ensures that all distances are observed through
the same solid angle.  The space density of RRLS is
calculated using a list of 334 \rrab stars spread over $\sim 330$ deg$^2$ of
the sky.

We followed \citet{sah85} for calculating the number density
of RRLS as a function of galactocentric distance $R_{gal}$.
In this method, the total number of objects $N$ found in a
solid angle $\omega$ is

\begin{equation}
N=\int{\omega \rho(r_\sun)\,r_{\sun}^2\,dr_\sun}
\end{equation}

Solving for the space density of RRLS, $\rho (r_\sun)$,

\begin{equation}
\rho(r_\sun) = \frac{1}{\omega r_{\sun}^2}\,\frac{dN}{dr_\sun}
\label{eq-rho}
\end{equation}

The quantity $dN\! /dr_\sun$ is estimated from a plot of the cumulative
number of RRLS versus $r_\sun$. First, all stars are sorted by increasing
$r_\sun$. Then, for each star $i$ at a distance $r_{\sun i}$, 
$\, dN\! /dr_\sun$ is the
local slope of the curve, which is calculated by fitting a straight
line to the 3 contiguous points centered at $r_{\sun i}$. Finally, distances $r_\sun$
are transformed to galactocentric distances $R_{gal}$ using equation
(~\ref{eq-rgal}). This way, it is possible to calculate a space
density at the $R_{gal}$ of each star (except for the ones at the
extreme $r_\sun$'s).

As noted by \citet{wet96b} this method works well for small solid
angles, where a heliocentric distance corresponds to a unique
galactocentric distance.  Our survey has a big solid angle which
covers a large range in galactic latitude and longitude, and Saha's
method will not work for such a region. \citeauthor{wet96b} proposed a
variation of this method which was applied to their driftscan survey,
similar in shape to ours, to calculate space densities based on a
model of a spherical halo or a ellipsoidal one.  We are
interested here in studying the density profile of the halo along different
lines of sight in order to see differences as a function of location
in the Galaxy.  Consequently, we divided our survey into small sub-regions,
where it was possible to use Saha's procedure.  The size of the
sub-regions must be large enough to contain a sufficient number of stars
for the calculations, but small enough to cover only a small range in
galactic coordinates.  We divided our $2\fdg
3$-wide strip in 19 pieces of equal area.  Each piece is 0.5 hr long in
right ascension for an area of 16.5 deg$^2$ (the area covered by the
gaps between CCDs has been subtracted).  The number of \rrab in each
of these slices of sky varies from 5 to 33.

\begin{figure}
\plottwo{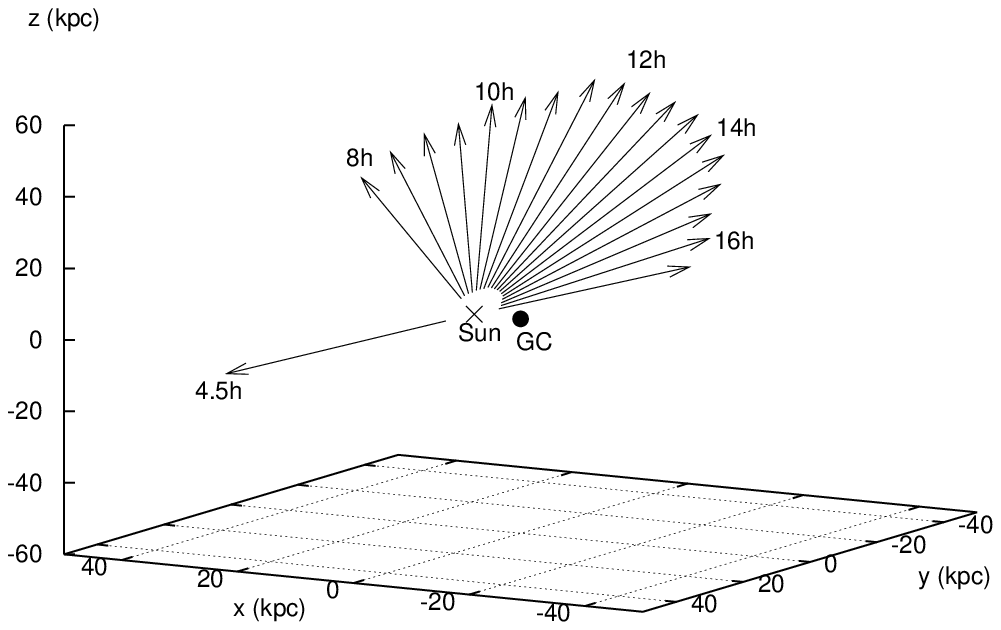}{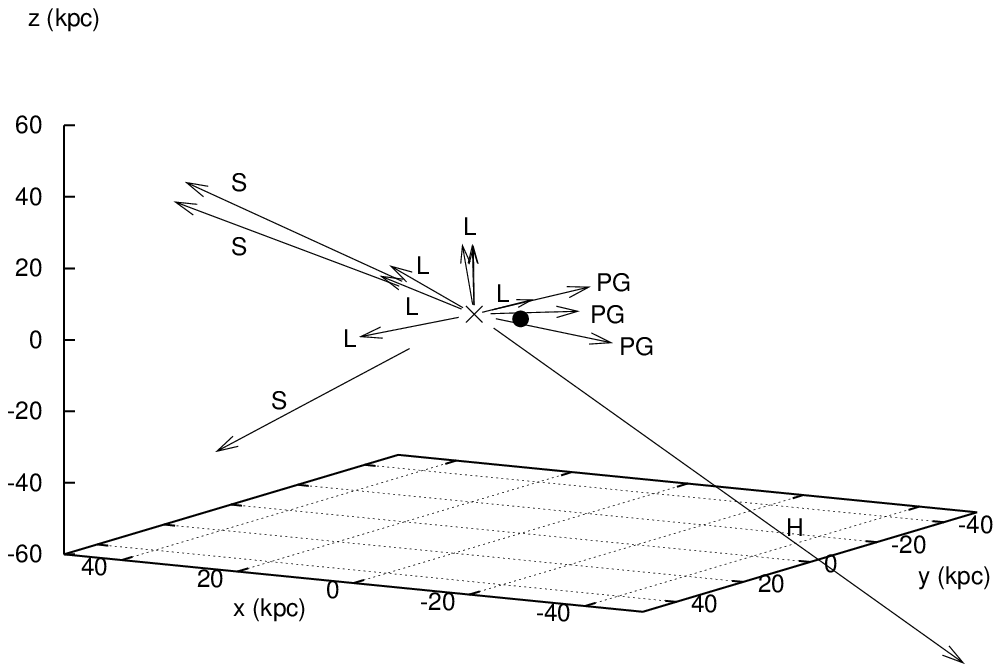}
\caption{Left: Line of sights of each sub-region in the QUEST survey in
galactic coordinates. The labels indicate right ascension.
Right: Other surveys for RRLS in the halo:
Lick (L), Palomar-Groningen (PG), Saha's (S) and Hawkins (H)}
\label{fig-los}
\end{figure}

Figure~\ref{fig-los} shows the line of sights of each of our
sub-regions on the sky. The length of the arrows indicate the range in
distance from the Sun covered in each sub-region which was calculated
from the limits in V magnitude of the survey and the extinction in
that particular direction.  Table~\ref{tab-regions} has the
coordinates and range of $R_{gal}$ covered in each sub-region.  For
comparison Figure~\ref{fig-los} also shows, at the same scale, the
lines of sight and depths of previous surveys of RRLS aimed to
calculate the space density of the halo: the Lick survey
\citep{kin65,kin66,kin82}, the Palomar-Groningen survey
\citep{pla66,pla68,pla71}, Saha's survey \citep{sah85} and Hawkins'
survey \citep{haw84}. The area covered by each line of sight in those
surveys (all photographic) varies from 16 to 43.6 deg$^2$ and each one
has different degrees of completeness. The details of each survey are
summarized in \citet{wet96b} (see also Table 4-2 in \citealt{smi95}).
One of the Lick's fields, MWF 361 or RR I, which has been extensively
studied \citep{kin65,kin85,pre91}, has partial overlap with our
sub-region $\alpha =16\fh 0 - 16\fh 5$.

Because of its special shape and characteristics, Figure~\ref{fig-los}
does not include the CTI survey of \citet{wet96a}.  The survey is
similar to ours in the sense that it is a driftscan survey. It covers
a very narrow strip ($8.\!' 3$) over all right ascensions for a total
area of 36 deg$^2$, up to a distance $r_\sun$ of 30 kpc.  Also, we do not
show the RR Lyrae candidate searchs made with data from the
SDSS \citep{ive00,ive05} because their completeness are relatively low.
The line of sights searched by \citet{ive00} have
similar direction than ours (but longer) in the range of $\alpha=
10\fh 7 - 15\fh 8$.

Figure~\ref{fig-los} shows that our survey covers a wide area of the
outer parts of the halo where few RRLS have been discovered.  Of the
previous surveys that identified RRLS by light curve and period,
only Saha's and Hawkins' surveys ventured before in the $r_\sun>30$ kpc
region of the halo.  Notice that previous surveys were mostly made
toward 3 directions: the galactic center, the galactic anti-center and
the north galactic pole. Our survey provides a homogeneous sample of
RRLS in a wide range of galactic coordinates.

In order to correct for incompleteness, the space density given by
equation~\ref{eq-rho} was divided by the completeness of the survey in
that part of the sky.  The completeness for the \rrab in the QUEST
survey is high ($>80\%$) over most of the region but decreases toward
the ends of the strip.  The completeness was calculated using
extensive simulations as a function of $\alpha$, separately for bright
($V<18.5$) and faint stars, and it is shown in Figure 10 of Paper I.
Small corrections have been made to these completeness estimates because 
we removed the nonvariables mentioned in secton 2

The number density of RRLS as a function of $R_{gal}$ along the
different lines of sight is shown in Figure~\ref{fig-density}. We have
plotted also the space densities of RRLS in Lick field ``RR I'', as
determined by \citet{pre91}, in the panel of $\alpha=16\fh 0-16\fh 5$. The
fact that we get similar results in the range of distances of overlap
give us confidence that our completeness estimates are
correct. \citet{pre91} used a different method to estimate densities,
which basically consists of calculating the volume occupied by a fixed
number of stars of continuously increasing $r_\sun$.

The most striking feature of the density profiles of Figure~\ref{fig-density}
is the big over-density of RRLS at $R_{gal}>40$ kpc in the regions 
$\alpha=13\fh 0 -15\fh 5$, which is due to the Sgr tidal tail (see also
Figure~\ref{fig-sky}).

\section{SMOOTH DENSITY CONTOURS}

Traditionally, the halo has been pictured as a region where the
density contours smoothly vary with $R_{gal}$ as a power law:

\begin{equation}
\rho (R_{gal}) = \rho_0 (R_{gal}/\!R_0)^n
\end{equation}

\noindent
where $\rho_0$ is the local space density of RRLS. The exponent $n$
has been determined by \citet{wet96b} by combining the results of
several surveys of RRLS and has a value of $n=-3.02\pm 0.08$. They
fitted the power law using data with $R_{gal}$ from 0.6 to 80 kpc,
with the caveat that only 9 objects with $R_{gal}>30$ kpc were
available.  \citet{amr01} have determined a value of
$\rho_0=4.6\pm0.4$ kpc$^{-3}$ with \rrab stars from the first results
of the ROTSE all-sky survey.

\begin{figure}
\plotone{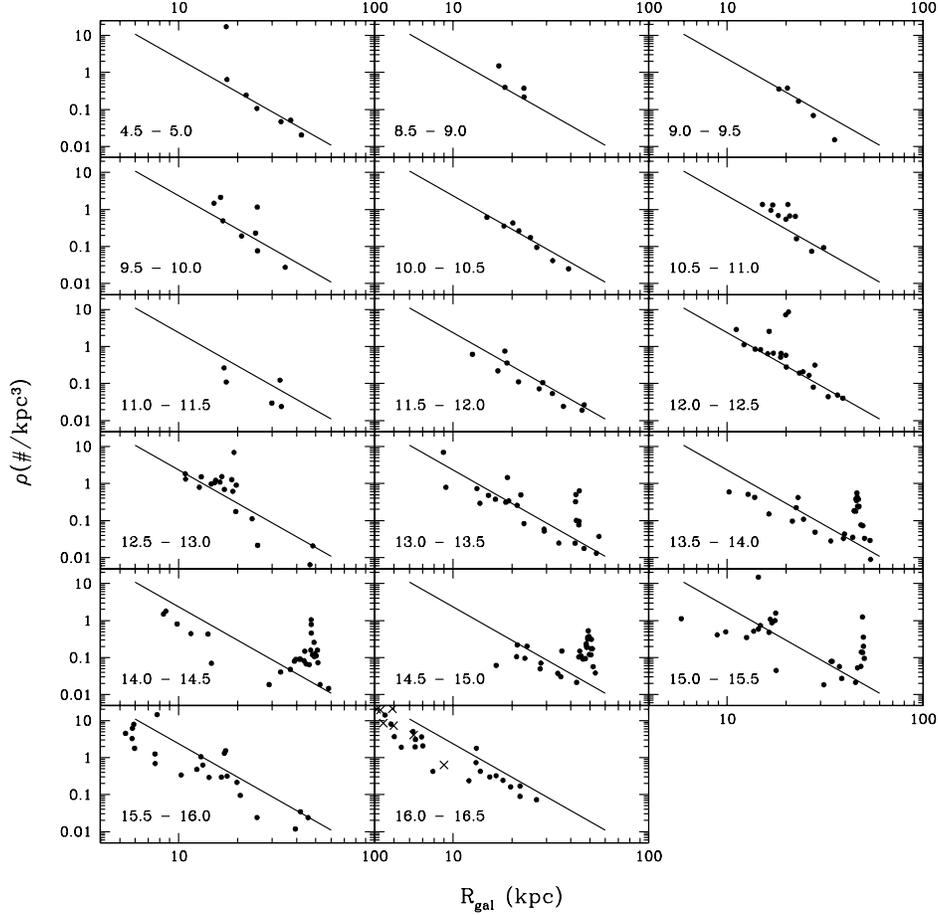}
\caption{Number of RRLS per unit volume as a function of
galactocentric distance. Each panel represents a different line of sight
which covers the right ascension range indicated in the lower left corners.
The $\times$'s in the panel corresponding to 
$\alpha=16\fh 0 -16\fh 5$ are the space densities
calculated by \citet{pre91} for Lick's field RR I, which partially
overlaps that sub-region. The lines show a power law with $\rho_0=4.6$ 
kpc$^{-3}$ and $n=-3.0$.}
\label{fig-density}
\end{figure}

The power law with the above values is shown in the density profiles
of Figure~\ref{fig-density} as a reference.  In principle, we do not
expect a perfect fit of the data to a power law not only because of
statistical fluctuations but also because the outer halo may be filled
with sub-structures, as predicted by hierarchical models of galaxy
formation (eg. \citealt{bul01}).  In fact there are several large
regions of unexpectly high density in our survey, which are seen 
in Figure~\ref{fig-density} as
large departures from the solid lines at certain distances $R_{gal}$.
Ignoring these over-densities, our first model, a spherical halo with
a power-law density fall-off, does a fair job describing the density
profiles along the different lines of sight, except in the regions
with $\alpha>14$ h for which it over-estimates the space density of
RRLS. As seen in Figure~\ref{fig-los} and Table~\ref{tab-regions}
these sub-regions have lines of sights approaching the galactic center
and thus, they contain several objects at small $R_{gal}$. Since this
power law reproduces well the space densities at larger $R_{gal}$,
there is either a problem with the parameters of the power law or the
assumption of an spherical halo.

Results from previous surveys toward the galactic center direction led
several groups to propose density contours that are flatter in the
inner halo \citep{kin65,wes87,haw84,kin94}.  \citet{pre91}
constructed an empirical model of a halo with variable flattening
which fit simultaneously RRLS density profiles toward the north
galactic pole and the GC. The model consists of isodensity contours
described by ellipsoids of revolution of semi-major axis $a$, whose
flattening $(c/\!a)$ is given by

\begin{equation}
c/\!a = \left\{ \begin{array}{ll}
                (c/\!a)_0+[1-(c/\!a)_0](a/\!a_u) & \mbox{if $a<a_u$} \\
                1 & \mbox{if $a>a_u$}
                \end{array}
        \right.
\label{eq-ca}
\end{equation}

\noindent
with $(c/\!a)_0=0.5$ and $a_u=20$ kpc. Consequently, at large
distances from the GC, this model has spherical density contours.

Our sample of RRLS provides a test of models with different halo
shapes and whether they can simultaneously fit the data along all lines
of sight.  We tested 3 different models of the halo: a spherical one
(shown in Figure~\ref{fig-density}), a flattened halo with constant
$(c/\!a)=0.6$, and \citet{pre91}'s model with variable flattening.

In the non-spherical models, the density contours are described by

\begin{equation}
\rho(a)= \rho_0 (a/\!R_0)^n
\end{equation}

\noindent
where

\begin{equation}
a=\sqrt{x^2 + y^2 + \left(\frac{z}{(c/\!a)}\right)^2},
\end{equation}

\begin{figure}
\plotone{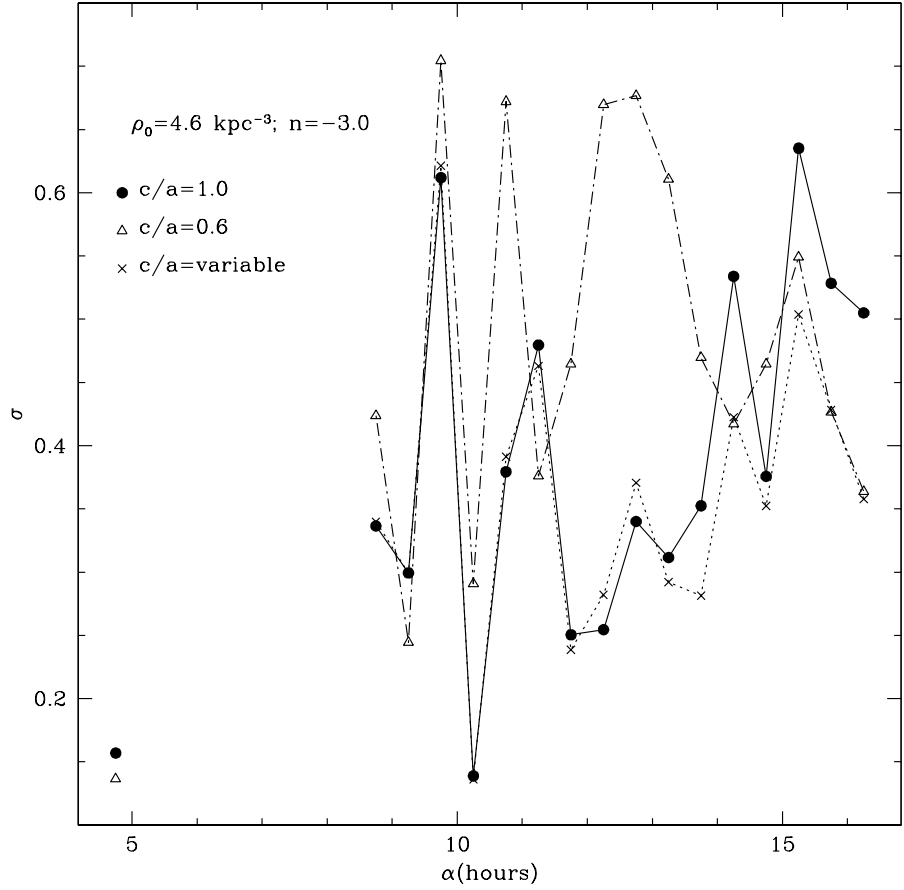}
\caption{Standard deviation of the data points to a density power law
with $\rho_0=4.6$ kpc$^{-3}$ and $n=-3.0$ for the
3 different models discussed in the text: a spherical halo (solid circles),
flattened halo with c/a=0.6 (open triangles) and
variable flattening ($\times$'s).}
\label{fig-model}
\end{figure}

\begin{figure}
\plotone{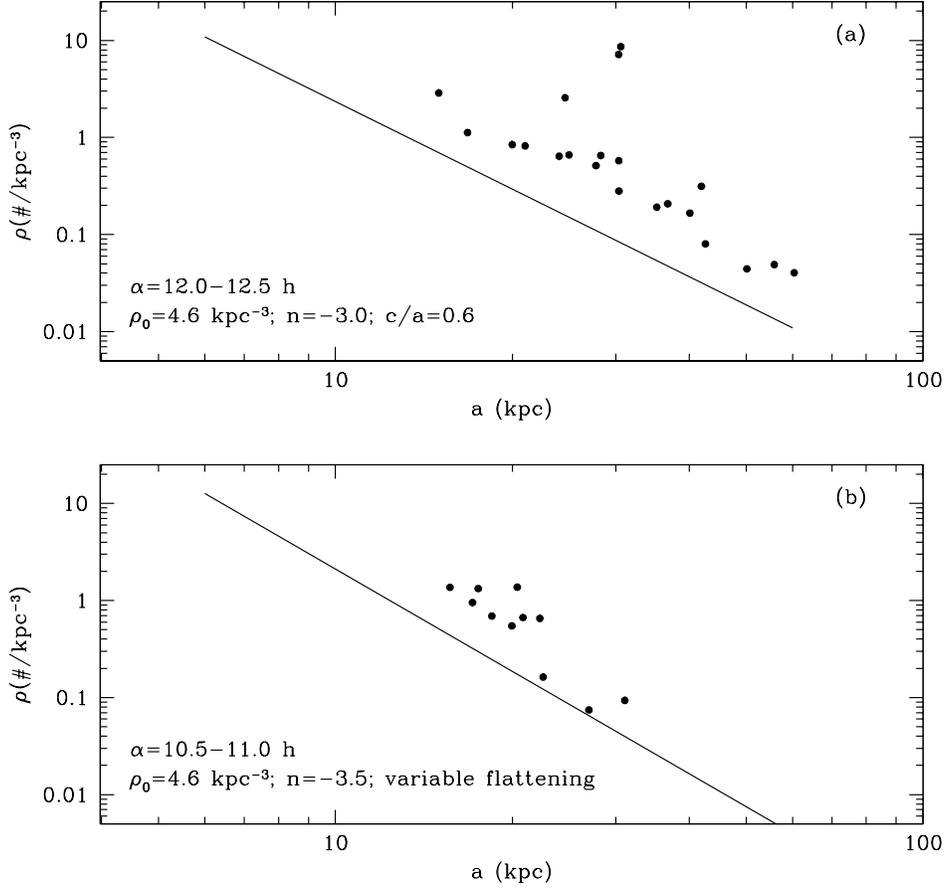}
\caption{Examples of models that do not fit the observed space density
of RRLS
(a) A flattened halo ($(c/\!a)=0.6$) underestimates the density
$\rho$ at high galactic latitudes, for example
in the panel corresponding to $\alpha=12\fh 0-12\fh 5$ ($b\sim 60^\circ$).
(b) A steeper power law ($n=-3.5$) does not reproduce the density profile at
$\alpha=10\fh 5-11\fh 0$, not even with the variable flattening model}
\label{fig-badmodel}
\end{figure}

\begin{figure}
\plotone{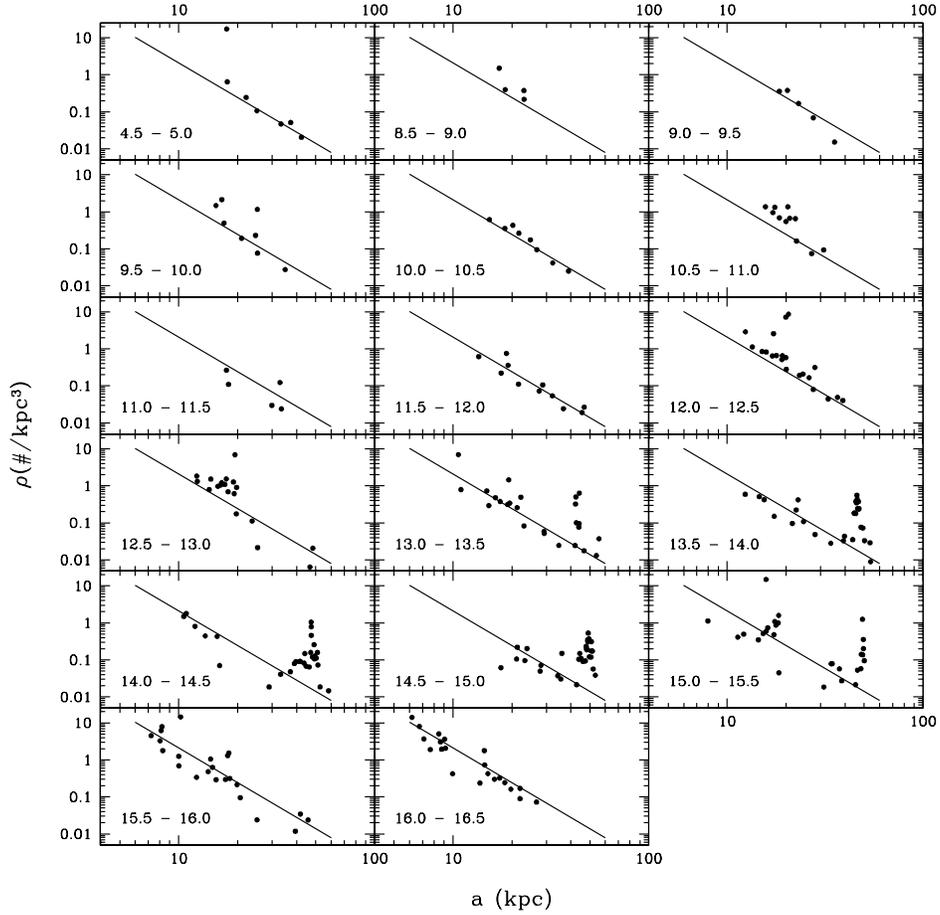}
\caption{Same as Figure~\ref{fig-density} but for the model of
variable flattening of the halo. Notice that the $x$-axis is now
$a$, the semi-major axis of the ellipsoids, which is equal to $R_{gal}$ for
$a\ge 20$ kpc.}
\label{fig-newdensity}
\end{figure}

To see which model best describes our data, we calculated the standard
deviation of the data points about the power law in each sub-region.
For these calculations we left out the the most obvious over-densities
of RRLS in Figure~\ref{fig-density}. Specifically, we did not include any
star with $R_{gal}> 40$ kpc in the range $13\fh 0 < \alpha < 15\fh 4$,
because they are known to belong to the Sgr stream. We also took out
some other stars causing notable over-densities:
3 stars at $R_{gal}\sim 18$ and $4\fh 5 < \alpha < 5\fh 0$;
5 stars at $R_{gal}\sim 20$ and $12\fh 0 < \alpha < 13\fh 0$;
and the 5 known variables of the globular cluster Pal 5 at
$R_{gal}\sim 17$ and $15\fh 0 < \alpha < 15\fh 5$.
Except for the Sgr and Pal 5 stars, the elimination of the other 
over-densities made no significant differences in the results we
describe here.
With the parameters $\rho_0=4.6$ kpc$^{-3}$ and $\,n=-3.0$ the three 
models produce the dispersions that are plotted as
functions of right ascension in Figure~\ref{fig-model}.  As expected
from Figure~\ref{fig-density}, the spherical halo (filled circles)
produces large deviations at $\alpha>14$h. On the other hand, a halo
with large, constant flattening (open triangles) yields a good fit of
the data in the regions toward the GC ($\alpha>15\fh 5$), 
but it does not reproduce the observed density
profile at high galactic latitudes ($9\fh 5 < \alpha < 14\fh 0$, see
Figure~\ref{fig-badmodel}a).  In those regions, the power-law tends to
underestimate the observed density of RRLS.
The model with varying flattening yields the best fit since it produces 
small and similar standard deviations over all lines of sight (see Figure~\ref{fig-newdensity}).
In particular, this model provides a better description of the data 
than the spherical model at $\alpha > 14h$ (cf. Figures~\ref{fig-density} and
\ref{fig-newdensity}).

The values of $n$ and $\rho_0$ used so far were both determined by
assuming a spherical halo, and they may not be optimal for the other
models.  For example, \citet{wet96b} found a steeper power law
($n=-3.53\pm 0.08$) when the variable flattening model was applied to
their data.  Also, a slightly higher value of the local density
($\rho_0=5.8\pm0.7$ kpc$^{-3}$) is found by counting RRLS very close
to the Sun, instead of averaging the value of the density at
$R_{gal}=R_0$ in several directions of the Galaxy \citep{amr01}.
Therefore, we repeated the experiment with these other plausible
density power laws and the best result was always found with the
variable flattening model.  For example, a steeper power law
($n=-3.5$) does not describe well the region of $\alpha=10\fh 5 - 11\fh 0$ with
any model (Figure~\ref{fig-badmodel}b) unless the local density takes
a much higher value, which is then unreasonable for all other lines
of sight and inconsistent with published measurements of that quantity.

We used the data over all the regions to derive the best parameters
for a power law with all the tested models. The densities along all
lines of sight were averaged in bins of equal size in $\log a$ (or
$\log R_{gal}$ in the case of the spherical model) and fit by the method of
least squares.  The results are presented in Table~\ref{tab-fit}. 
On average, both the flattened halo model and the variable flattening one
produced the best overall fit to our data (lowest rms). 
However, we favored the
variable flattening model because it best reproduces both the 
local density found by \citet{amr01}, and the slope $n$ 
from previous works, including other tracers ($-3.5$ for Globular Clusters
\citep{zin85} and BHB \citep{pre91}).  
Notice that 52 stars with $R_{gal}>30$ kpc were used in the
fit, $\sim$6 times more than the number used in previous studies
\citep{wet96b}.

\begin{figure}
\plotone{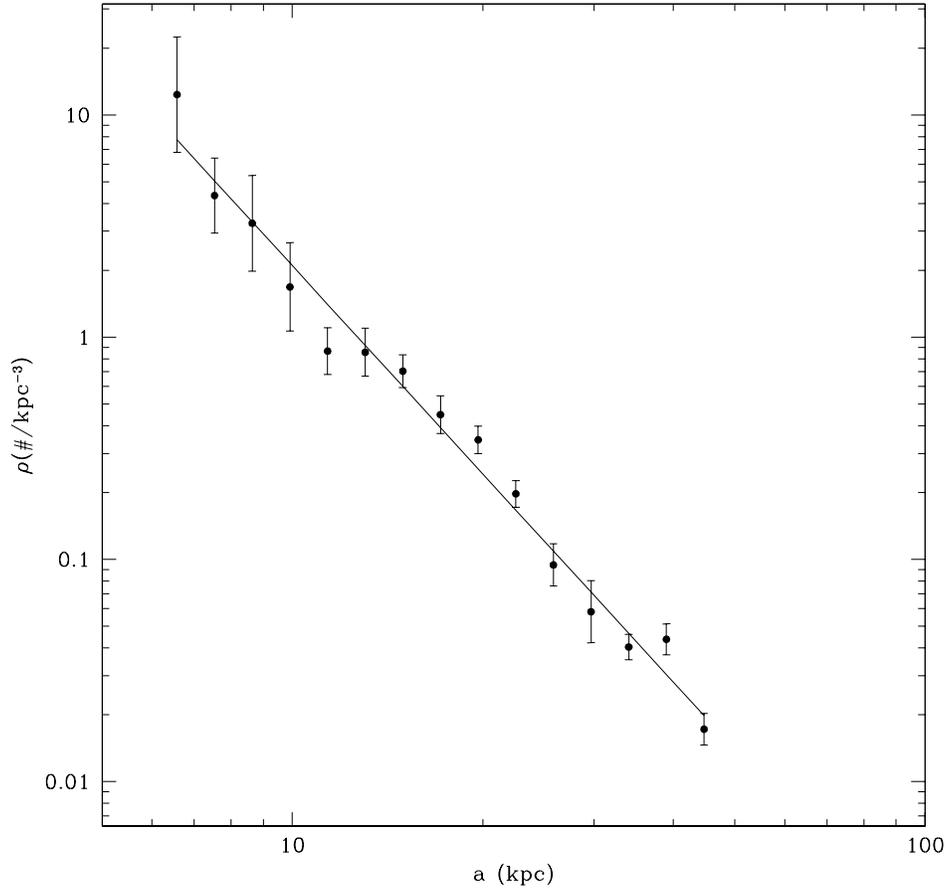}
\caption{Averaged space density profile of RRLS using the variable flattening
model for the halo. The solid line is the best fitted power law with $n=-3.1$
and $\rho_0=4.2$ kpc$^{-3}$. Error bars are the standard deviations of the mean
density in each bin.}
\label{fig-avden}
\end{figure}

In Figure~\ref{fig-avden}, the densities averaged over all lines of sight are compared
with the best fitting profile of the variable flattening model.  The fit
is reasonable, but with a larger than expected number of points deviating
by more than one standard deviation. 
This may be due to the failure of the model to account for the lumpy 
nature of the halo and the variations in the specific frequency of RRLS.
Nonetheless, this model serves our purpose of providing a 
description of the "background" of RRLS.

There is no evidence for a steeper power law ($n\sim -5$) beyond 25
kpc as was suggested by \citet{sah85}. The underdensity that he
observed in one of his fields may be simply due to the clumpy nature
of the outer halo. This might also be the explanation 
for the observation by Ivezic et al (2000) of an edge to the halo at 50-60 
kpc (see Ivezic et al. 2004).

\section{IDENTIFICATION OF OVERDENSITIES}

We have performed the following analysis to identify overly dense regions 
that may escape detection by visually examining Figure~\ref{fig-sky}.  
Many of these 
feature need to be confirmed by radial velocity and/or proper motion 
measurements.  We have restricted this analysis to stars with 
$\langle V_0 \rangle > 14.8$ because at brighter magnitudes the differences in the 
saturation of limits of the CCDs in the QUEST camera produce some variation 
in completeness.

The first step in our analysis was the development of a 
computer code to recognize groups of RRLs that may be 
statistically significant.  For each star ($i$) in the QUEST 
catalogue, the code computed the distances ($d_{ij}$) from $i$ to 
each of the other stars $j$ and kept as neighbors the stars within a 
distance limit ($d_{limit}$).  It then created a ranking of the $d_{ij}$ values 
of the neighbors and computed the volumes centered on star $i$ that 
included in turn each of its more distant neighbors.  Because the 
QUEST survey is limited in $\delta$ to $2.3\degr$, these volumes 
are non-spherical, and we approximated them with cylinders with 
heights in the $\delta$ direction.   Each volume contains star $i$, 
the star $j$ at distance $d_{ij}$ and all stars with smaller distances 
from star $i$, if any.     
The number of stars observed in a volume, $n_{obs}$, was then compared 
with the number expected in the absence of an overdensity.  Our model for 
the density contours (see above) provides the number density of \rrab 
variables at the position of star $i$.  These densities were then multiplied 
by factors that take into account the incompleteness of survey in type 
ab and in type c variables, which vary with $\alpha$ and $\langle V \rangle$ 
(Paper I).   
The resulting number density for all types of RRLS was then multiplied 
by the volume to yield $n_{exp}$ , the expected number of RRLS, which 
was rounded to the nearest integer. We then computed the probability, $P$, 
that $N \ge n_{obs}$, given that $N$ is a random number following 
a Poisson distribution of mean $n_{exp}$, 

\begin{equation}
P(n_{obs} , \infty, n_{exp}) = 1 - \sum_{x=0}^{x=n_{obs}-1} P(x, n_{exp})
\end{equation}
  
Star $i$ and each of its neighbors that occupy a volume where $P \le P_{limit}$ 
are considered members of a group.  If none of these stars has been previously 
assigned to a group,  then a new group number is assigned.  Otherwise, they 
are considered additional members of the previously assigned group.   In addition 
to group number, star $i$ is tagged by the lowest $P \le P_{limit}$ 
that occurred in the calculations of the values of $P$ for the small volumes. 

The free parameters in this analysis are the choices for $d_{limit}$ and $P_{limit}$.  
For $d_{limit}$, we chose $0.18\, r_\sun$, which is approximately $2.5\, \sigma_{r_\sun}$.  
Consequently, two stars that lie at the same distance and direction have a 
probability of $\sim 0.99$ of being considered neighbors in this analysis.  Increasing 
$d_{limit}$  significantly above this value produces overlap between what may be separate 
over-densities.   Decreasing $d_{limit}$ significantly may mean that some 
small over-densities 
are missed because of distance errors.  
For $P_{limit}$, we chose the value $1.0 \times 10^{-3}$, 
which at first glance may appear smaller than necessary.  Because there are 457 stars 
in the revised QUEST catalogue and a minimum of a few stars within $d_{limit}$ of each,  
thousands of calculations of $P$ are made.  It is fairly common for random fluctuations 
alone to produce a few small groups with $P \le 1 \times 10^{-3}$, as our Monte Carlo 
simulations indicate (see below). 

In Figure~\ref{fig-overdensities}, 
we have used different symbols to indicate the confidence that we attach 
to a star's membership in an overdensity.  Values of $P$ in the ranges  
$1 \times 10^{-3} \ge P > 1 \times 10^{-4}$ , $1 \times 10^{-4} \ge P > 1\times 10^{-5}$ , 
and $P \le 1\times 10^{-5}$ indicate with low, medium, and high confidence, respectively, 
that a star is considered part of an overdensity.   We must emphasize that some of the 
stars identified as belonging to a group will turn out to be clearly nonmembers once their 
motions are measured because every overdensity is overlaid on a "background" of 
unrelated RRLS.    
We are therefore identifying stars that are candidate members of the groups, and their $P$ 
values provide an estimate of relative likelihood of membership.  The sparse globular cluster 
Pal 5 lies within the survey region (see Figure~\ref{fig-sky}).  
Not surprisingly, its five RRLs, which were 
all detected by the QUEST survey, produce a very significant overdensity which is 
not of interest 
because the cluster has been known for decades.  We therefore removed these 5 stars from the 
survey before performing the search for overdensities.   One of the identified overdensities 
is in close proximity to Pal 5 and is undoubtedly related (see below).   For each overdensity 
recognized by the above procedure, the quantity $P_{ave}$ was calculated by averaging the $P$ 
values that were assigned each member of the group.  The above limits on $P$ have also been 
applied to $P_{ave}$ in order to identify groups of low, medium, and high confidence.  
The groups that seem most likely to be real and their confidence category are listed in 
Table~\ref{tab-overden}, 
where also are listed the position of the group in Figure~\ref{fig-overdensities}, 
the number of members ($n_{obs}$), $P_{ave}$,  
and the frequency ($F$) with which a similar group occurred by chance in the Monte Carlo 
calculations. $F$ is equal to the total number of groups with $N\ge n_{obs}$  and average 
$P \le P_{ave}$ that were produced by the simulated surveys divided by the total number 
of artificial surveys ($1\times 10^4$).

In the Monte Carlo simulations, the volume of the galactic halo covered by the  QUEST survey 
was subdivided by $\alpha$ into intervals of 0.25 hours ($3\fdg 75$) and by 
$\langle V_0 \rangle$ into intervals of 0.25 mag.  
No subdivision was made by $\delta$.  The position in the 
Galaxy of the center of each subdivision was computed using $M_V = +0.55$ 
for RRLS,  and the number 
density of \rrab at this position was calculated from our model for the 
background of type 
ab RRLS.  
This number density was corrected for the incompleteness of the QUEST survey using the 
same factors that were used with the real survey.  
It was then multiplied by the volume delineated 
by the intervals in $\alpha$, $\delta$, and $\langle V_0 \rangle$.  
After rounding to the nearest integer, 
this yielded $n_{exp}$ for the volume.  
To provide for statistical variation, we computed a 
random number ($N$) from a Poisson distribution (a Poisson deviate) 
with mean equal to $n_{exp}$, 
which employed a random number as a seed variable.   
These $N$ artificial stars were then randomly 
distributed in $\alpha$,  $\delta$, and $\langle V_0 \rangle$ 
within the limits of the subdivision using 
random a number generator with different seed values for each quantity.  The calculation then 
proceeded to the next subdivision where $n_{exp}$ was again calculated for the new position and 
different random selections were made for the seed variables in the calculation of the Poisson 
deviate and in the distribution of the stars in position and in magnitude.  These calculations 
were continued until the area of the QUEST survey  was covered.   The program for identifying 
overdensities that we described above was then used to find and characterize any groups that 
were 
produced in this artificial survey.  

In order to see what size groups and their 
$P_{ave}$ values that 
might have arisen by chance in the QUEST survey, a total of $1\times 10^4$ artificial 
surveys were 
produced by starting each Monte Carlo calculation with a different seed value.  
Each of these surveys 
was fed to the program for finding overdensities and the values of the frequencies 
$F$ (see above) 
were computed.  
Experiments with different sizes for the intervals in $\alpha$ and $\langle V_0 \rangle$ showed 
that these Monte Carlo calculations were not particularly sensitive to their choices as long as 
the intervals were not so large to encompass a significant variation in number density and not 
so small that $n_{exp} = 0$ for many subdivisions.   One of these artificial surveys, which is 
typical in terms of the number and sizes of overdensities, is displayed in 
Figure~\ref{fig-mc}.  The artificial stars that are in overdensities 
are labeled in exactly the same manner as the stars in 
RRLS in the real survey.  
Since the same model for the smoothed density contours was used both to 
generate the artificial survey and to identify the overdensities, the ones that appear in the 
artificial surveys are entirely due to statistical fluctuations.  Once can see from this figure 
that random variations alone can produce significant overdensities, even ones that meet our 
criteria for the high confidence category.    Not surprisingly, the majority of the randomly 
occurring overdensities contain very few stars.

\section {DISCUSSION}

Table~\ref{tab-overden} lists the 6 most significant overdensities in the 
QUEST survey according to our group 
finding routine.  With the exception of Group 2, we will discuss each of these groups in order 
of their sizes.  Group 2 lies in the region of the survey where the interstellar extinction is 
high and variable.  Because our estimate of the completeness of the survey in this region is 
suspect, we consider Group 2 to be only a marginal detection  of halo substructure.

\subsection {Group 5, The Sagittarius Stream}

The largest and by far the most significant over-density is Group 5, at a 
mean distance of 50 kpc, which was produced by 
the tidal destruction of the Sgr dSph galaxy.  According to our 
group finding routine, 
Group 5 contains 104 candidate members, which is $23\%$ of the whole sample of QUEST RRLS.  
Its 
$F$ value (see Table~\ref{tab-overden}) 
indicates that none of the $10^4$ Monte Carlo simulations produced a group of 
this size or larger with $P<P_{ave}$ .   
 
This striking overdensity was discovered in the spatial distribution of A stars 
\citep{yan00} and 
RRLS candidates \citep{ive00}.  Shortly thereafter, a main-sequence was found in the 
color-magnitude 
diagram of a small region \citep{m-d01}.  It has also been detected in the distributions of 
carbon stars \citep{iba01}, F type main-sequence stars \citep{new02}, 
and M giants \citep{maj03}.   
Our initial results from the QUEST survey \citep{viv01} provided additional evidence for a 
large overdensity of RRLS.   The distributions of these stars and the measurements of radial 
velocities for subsamples \citep[e.g.][]{doh01,maj04,viv05}
provide conclusive proof that this feature is part of the leading stream from the 
Sgr dSph galaxy.   
Figures~\ref{fig-sky} and \ref{fig-overdensities}  
show that there are very few QUEST RRLS at large $r_\sun$ that are not part of Group 5.  
This observation and similar ones using the different types of stars mentioned above illustrate 
that the debris from the Sgr dSph galaxy is a major contributor to the outer galactic halo. 

\subsection {Group 4, the ``12.4 h clump'', the ``Virgo Stellar Stream'' 
and the ``Virgo Overdensity''}

Group 4, which after the Sgr Stream is the most prominent feature in 
Figures~\ref{fig-sky} and \ref{fig-overdensities}, contains 42 
potential members.  The $F$ value of this group (Table~\ref{tab-overden}) 
indicates that a 
few of the Monte Carlo 
simulations produced artificial groups of equal or larger size and significance.   
Based on this 
information alone, there is a small, but not negligible probability, that this feature 
is an artifact.

The eastern edge of this feature was identified as an overdensity in our first report of 
QUEST results \citep{viv01}.  It became known as the ``12.4 h clump'' when its large size was 
recognized in the completed QUEST survey \citep{viv02,viv03}.
The survey 
of F type main-sequence stars in the SDSS \citep{new02} also revealed its large size.  
Figure~\ref{fig-overdensities} shows that it spans 
$12 \le r_\sun \le 20$ and $175 \le \alpha \le 205$ , which corresponds to a 
span of $\sim 9$ kpc at the mean of 17 kpc.  

The densest part of this feature is located at $\alpha \sim 186\degr$ and $r_\sun \sim 17$ kpc.  
Very recently, \citet{duf06} have shown that a subsample of the QUEST RRLS in 
this dense region and 
a subsample of the BHB stars in the same region of space from the 
\citet{sir04} survey have very similar radial velocities, indicating that they 
are part of a 
stellar stream.  The RRLS in the stellar stream have a 
low mean metallicity but a wide range in metallicity, 
which suggests 
that they are the debris from a low luminosity dSph galaxy. \citet{duf06} have 
suggested 
the name the Virgo Stellar Stream (VSS) for this feature of halo substructure, which they have 
traced over 106 deg$^2$ of the sky.  Until more radial velocity measurements are obtained, 
it is not 
clear that majority the stars of Group 4 belong to the VSS. \citet{duf06} noted 
that in 
addition to the VSS there is less significant evidence for two other moving groups 
containing smaller 
numbers of RRLS and  BHB stars, which may be a sign of that there is more to 
Group 4 than the VSS.

Group 4 also lies in the direction of the `` Virgo Overdensity'' that has been 
recently identified 
by \citet{jur05} in distribution of main-sequence stars in the latest photometry 
of the SDSS.   
They estimate that this feature covers $>1000$ deg$^2$ of the sky and lies in the range 
$5\le r_\sun \le 15$ kpc.  It therefore overlaps considerably with Group 4, and 
considering  distance 
errors of $\sim 20\%$ for the main-sequence stars, 
it also overlaps with the known members of the VSS, which lie at 
its distant edge.   
\citet{jur05} suggest that this feature is the tidal stream from a dwarf galaxy or 
possibly the 
galaxy itself.   While the exact relationships between the VSS, the Virgo Overdensity, 
and Group 4 are 
unclear at this moment, there is no doubt that at least part of Group 4 is a real 
halo substructure. 
 
\subsection{Groups 1 and 3, the Monoceros Stream?}

The next two most significant groups, Groups 1 and 3, may be related to each other.  
They lie at similar values of $r_\sun$ on either side of the region skipped by the 
QUEST survey because 
of the severe crowding of star images near the galactic plane.   The low $F$ values of 
these groups 
suggest that they are likely to be real.  
Group 1 is the less certain of the two because it lies in the 
range of $\alpha$ where it was difficult to estimate the completeness of the survey.  
Spectroscopic observations have been obtained for many of the stars in these groups and will be 
discussed in a later paper.

As we noted in a preliminary report of our results \citep{zin04}, Groups 1 and 3 lie 
approximately in the same direction and at the same distance as the Monoceros Stream that was 
discovered by \citet{new02} in the SDSS photometry and confirmed by the spectroscopic
observations of \citet{yan03}.  
This stream appears to be part of large-ring like structure that 
envelops the Milky Way \citep[see also][]{maj03,iba03,roc03,pen05,con05a,con05b}.
It is widely interpreted as the debris of tidally disrupted dwarf galaxy, but it is 
clearly not part 
of the Sgr Stream nor any of the other halo overdensities that we have discussed above.  
While the 
association of Groups 1 and 3 with the Mon Stream appears likely, this is not certain.   The 
measurements by \citet{kin04} of the radial velocities of a sample of RRLS in the 
anticenter 
direction did not reveal any that clearly belong to the Mon Stream.  

\subsection {Group 6, the overdensity near Pal 5}

As we noted above, we purposely removed the 5 RRLS in the globular cluster Pal 5 from the QUEST 
catalogue before searching for overdensities.    Group 6 lies very close to Pal 5 and 
at exactly 
the same distance, which of course suggests a relationship with the cluster.   Consequently, we 
discount the relatively large $F$ value of Group 6, which suggests that the QUEST survey 
is expected 
to contain one or two groups of the size and significance of Group 6 that are nothing more than 
statistical fluctuations. 

In the top diagram of Figure~\ref{fig-pal5}, 
we have plotted the 5 previously identified RRLS in Pal 5, the 6 
stars in Group 6, and other QUEST RRLS in the region.   The bottom diagram of 
Figure~\ref{fig-pal5} shows the 
distribution of the same stars on the sky along with a rough outline of Pal 5 and its 
tidal tails, 
as revealed by the analysis of SDSS photometry by \citet{ode01,ode03}.   
One can see from the upper diagram that all 6 stars in Group 6 lie at the same $r_\sun$ 
as Pal 5 to within the errors.  Two of the Group 6 stars, RRLS 403 and 405, 
lie approximately 9.2 and 11.2 arcmin., respectively, from the cluster center 
(see lower diagram).  The extensive photometry of \citet{ode02} shows 
that there are many Pal 5 stars at similar angular distances; consequently the 
association of these RRLS with the cluster is nearly certain.  Another RRLS (393),  
which was not 
identified as member of Group 6 because it lies too far away from the group's center 
is also likely 
to have been once a member of Pal 5.  Its position on the sky and distance are 
consistent with it 
being part of the southern tidal stream from Pal 5.  It is interesting to note 
that Pal 5 has been 
considered odd among globular clusters because all five of its previous known 
RRLS are type c.   
Two of the three probable members identified here are type ab, which partially 
removes the imbalance in 
the numbers of these types of variables.

It is less certain that any of the other members of Group 6 are related to Pal 5, for they are 
offset from the cluster to the east by significant amounts.   However, as reported in 
\citet{zin04}, our preliminary measurements suggest that the two members of Group 6 
near $\alpha \sim 230\degr$ have radial velocities and metallicities that are consistent 
with membership in the cluster.  This possibility is being explored as the 
measurements are refined.

\subsection{Other Overdensities}

One can see from Figure~\ref{tab-overden} 
that the group finding routine has identified a number of 
other small groups.
While the $F$ values of these groups suggest that they may be not be real, this 
cannot be discounted 
entirely until their motions are measured.  \citet{kin04} have shown that 
small groups of 
co-moving RRLS exist in the halo, which they suggest may be the remains of disrupted globular 
clusters.  Our spectroscopy of few of the small groups in the QUEST survey 
will be reported later.

\section{Summary}

Our major result is that the Galactic halo contains substructures over a spectrum of sizes.  
While much remains to be learned about the features identified here and by other surveys, there 
is strong evidence that at least some were produced by the 
merger of smaller galaxies with the Milky Way.
These merger events have clearly deposited a large number of stars in the halo, as predicted by 
models of the hierarchical picture of galaxy formation \citep[e.g.][]{bul05}.
While this may seem so obvious now, it represents a large departure 
from the traditional picture of the galactic 
halo that could be characterized by smooth density contours and a power-law drop-off.  

\acknowledgments

This research project
was partially supported by the National Science Foundation under grant
AST-0098428 and AST 0507364. 
AKV thanks Bruce Carney for several interesting and helpful comments
and suggestions. The QUEST survey for RR Lyrae stars uses the facilities of
the Observatorio Nacional de Llano del Hato, which is operated by CIDA 
for the Ministerio de Ciencia y Tecnolog{\'\i}a, Venezuela. We thank the anonymous
referee for encouraging us to discuss about the specific frequency of RRLS in
different populations.

\begin{figure}
\plotone{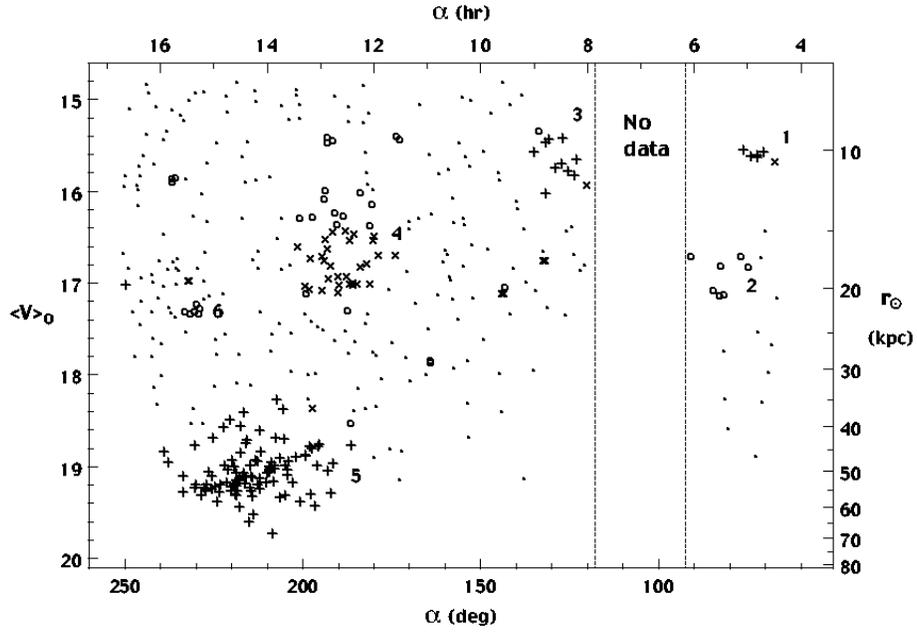}
\caption{The values of mean V magnitude, corrected for extinction, of the QUEST RRLS are 
plotted against right ascension.  The 5 RRLS in the globular cluster Pal 5 have been removed 
from this sample.  The crosses, $\times$'s, and open circles depict the RRLS that are assigned 
to groups with high, medium, and low confidence, respectively.  The small dots are RRLS that do 
not belong to groups according to our group finding routine.  The six most significant groups 
are numbered in order of increasing $\alpha$.}
\label{fig-overdensities}
\end{figure}

\begin{figure}
\plotone{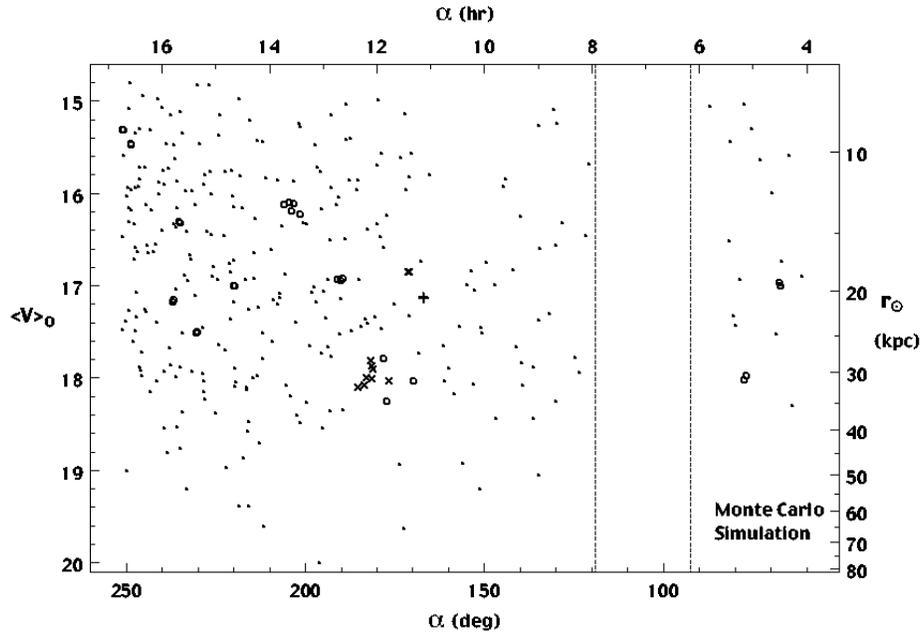}
\caption{For a typical Monte Carlo simulation of the QUEST RRLS survey, mean V magnitude, 
corrected for extinction, is plotted against right ascension.  The artificial stars depicted 
as crosses, $\times$'s, and open circles are assigned to groups with high, medium, and 
low confidence, respectively.  The small dots are artificial stars that are not assigned 
to groups.  The large group at $\alpha \sim 180$ and ${langle V \rangle}_0 \sim 18$ and 
all other clusterings in this diagram are due to statistical fluctuations.}
\label{fig-mc}
\end{figure}

\begin{figure}
\plotone{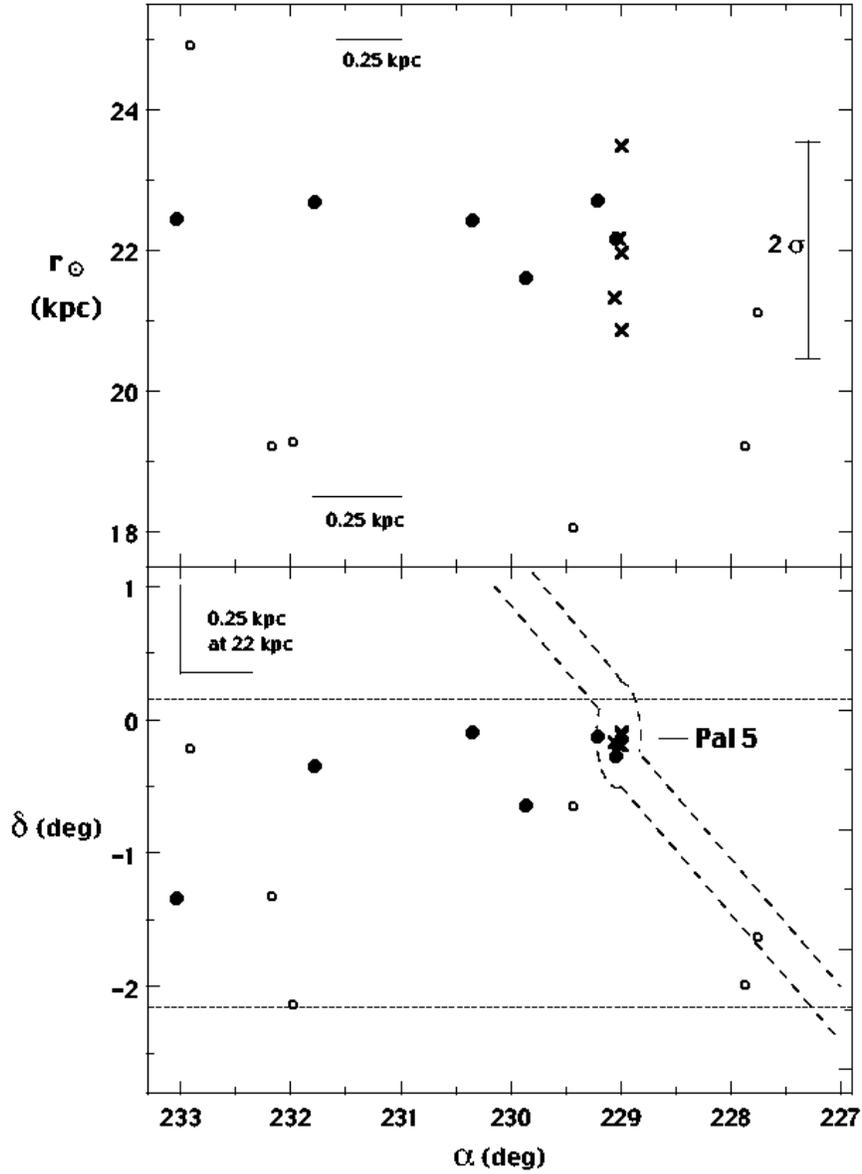}
\caption{These diagrams show the RRLS in the vicinity of the globular cluster Pal 5.  
The $\times$'s are the five previously known RRLS in Pal 5.  The QUEST values of 
$\langle V_0 \rangle$ were used in the calculation of their distances.  The solid 
circles are the RRLS belonging to Group 6.  The small open circles are other QUEST RRLS.  
The dashed horizontal lines in the lower diagram show the declination limits 
of the QUEST survey.  The other dashed lines delineate the boundaries of the tidal tails 
of Pal 5 \citep{ode03}. }
\label{fig-pal5}
\end{figure}

\clearpage
\begin{deluxetable}{lccrrc}
\tablecolumns{9}
\tablewidth{0pc}
\tablecaption{The Specific Frequencies of RR Lyrae Variables in Dwarf Spheroidal Galaxies}
\tablehead{
\colhead{Galaxy} & \colhead{$M_V$} & \colhead{[Fe/H]} & 
\colhead{$N_{obs}$\tablenotemark{a}} & \colhead{$N_{est}$\tablenotemark{b}} &
\colhead{$S_{RR}$\tablenotemark{c}} }
\startdata
Carina      & $-8.6 \pm 0.5$  & -2.0 & 75   & 76   & $28 \pm 6$ \\
Draco       & $-8.4\pm 0.2$ & -2.0 & 268  & 268  & $117\pm 12$ \\
Fornax      & $-13.0 \pm 0.3$ & -1.3 & 515  & 660  & $4.2\pm 0.2$ \\
Leo I       & $-11.5 \pm0.3$ & -1.5 & 74   & 74   & $1.9 \pm0.2$ \\
Leo II      & $-9.6\pm 0.3$  & -1.9 & 148  & 148  & $21 \pm2$ \\
Sagittarius & $-13.9\pm 0.5$ & -1.0 & 1700 & 5200 & $14.3\pm 0.2$ \\
Sculptor    & $-10.7\pm 0.5$ & -1.8 &	226  & 500  & $26 \pm1$ \\
Sextans     & $-9.2 \pm 0.5$   & -1.7 & 111  & 260  & $54\pm 7$ \\
Ursa Minor  & $-8.4\pm 0.5$  & -2.2 &	82   & 84   & $37 \pm9$ \\
And I       & $-11.8\pm 0.1$ & -1.5 &	99   & 582  & $11.1\pm 0.5$ \\
And II	    & $-12.6\pm 0.2$ & -1.5 & 	72   & 400  & $3.7 \pm0.2$ \\
And III	    & $-10.2\pm 0.3$ &	 -1.9 &  51   & 204  & $17.0\pm 1.3$ \\
And VI	    & $-11.5\pm 0.2$ &  -1.6 & 111  & 529  & $13.3\pm 0.6$ \\
\enddata	
\tablenotetext{a}{$N_{obs}$ is the number of RRLS discovered in the surveyed field}
\tablenotetext{b}{$N_{est}$ is an estimate of the total number of RRLS in the galaxy}
\tablenotetext{c}{$S_{RR}$ is the specific frequency of RRLS, which is the number of 
RRLS per unit $M_V$ , normalized to $M_V = -7.5$}
\tablerefs{Values of $N_{obs}$ taken from: Car: \citet{dal03}; Dra: \citet{kin02};
For: \citet{ber02}; Leo I: \citet{hel01}; Leo II: \citet{sie00}; Sgr: \citet{cse01};
Scl: \citet{kal95}; Sex: \citet{lee03}; UMi: \citet{nem88}; And I \& And III: \citet{pri05};
And II: \citet{pri04}; And VI: \citet{pri02}}
\label{tab-frequency}
\end{deluxetable}

\clearpage
\begin{deluxetable}{lrrcrrrrr}
\tabletypesize{\small}
\tablecolumns{9}
\tablewidth{0pc}
\tablecaption{Galactic coordinates and distances of the RRLS\tablenotemark{a}}
\tablehead{
\colhead{ID} & \colhead{$l$ ($^\circ$)} & \colhead{$b$ ($^\circ$)} &
\colhead{$V_0$} & \colhead{$x$ (kpc)} & \colhead{$y$ (kpc)} & \colhead{$z$ (kpc)} &
\colhead{$r_\sun$ (kpc)} & \colhead{$R_{gal}$ (kpc)}}
\startdata
217 & 299.05 &  61.87 & 16.23 &   4.9 &  -5.7 &  12.1 & 13.7 & 14.2 \\
218 & 299.38 &  60.74 & 14.53 &   6.5 &  -2.7 &   5.4 &  6.2 &  8.9 \\
219 & 299.42 &  61.98 & 14.79 &   6.4 &  -2.9 &   6.2 &  7.0 &  9.4 \\
220 & 300.00 &  60.95 & 18.96 &  -3.7 & -20.2 &  42.0 & 48.0 & 46.8 \\
221 & 300.19 &  61.79 & 16.43 &   4.4 &  -6.1 &  13.2 & 15.0 & 15.3 \\
\enddata
\tablenotetext{a}{The complete version of this table is in the electronic
edition of the Journal.  The printed edition contains only a sample.}
\label{tab-rrgc}
\end{deluxetable}

\clearpage
\begin{deluxetable}{crrc}
\tablecolumns{4}
\tablewidth{0pc}
\tablecaption{Coordinates of the sub-regions of the survey}
\tablehead{
\colhead{$\alpha$ (h)} & \colhead{$l$ ($^\circ$)} & \colhead{$b$ ($^\circ$)} & \colhead{$R_{gal}$ (kpc)} }
\startdata
4.5 - 5.0 & 198.2 & -28.3 & 15 - 70 \\
8.0 - 8.5 & 223.7 & 18.0 & 14 - 69 \\
8.5 - 9.0 & 227.7 & 24.5 & 14 - 70 \\
9.0 - 9.5 & 232.2 & 30.9 & 13 - 69 \\
9.5 - 10.0 & 237.3 & 37.1 & 13 - 66 \\
10.0 - 10.5 & 243.3 & 43.0 & 13 - 67 \\
10.5 - 11.0 & 250.5 & 48.6 & 12 - 65 \\
11.0 - 11.5 & 259.5 & 53.5 & 11 - 64 \\
11.5 - 12.0 & 270.7 & 57.6 & 11 - 66 \\
12.0 - 12.5 & 284.2 & 60.5 & 10 - 65 \\
12.5 - 13.0 & 299.5 & 61.8 & 10 - 63 \\
13.0 - 13.5 & 315.3 & 61.3 &  9 - 62 \\
13.5 - 14.0 & 329.7 & 59.1 &  8 - 62 \\
14.0 - 14.5 & 341.9 & 55.4 &  7 - 60 \\
14.5 - 15.0 & 351.7 & 50.8 &  7 - 59 \\
15.0 - 15.5 & 359.6 & 45.5 &  6 - 55 \\
15.5 - 16.0 &   6.1 & 39.7 &  5 - 51 \\
16.0 - 16.5 &  11.6 & 33.6 &  5 - 50 \\
16.5 - 17.0 &  16.3 & 27.3 &  4 - 45 \\
\enddata
\label{tab-regions}
\end{deluxetable}

\clearpage
\begin{deluxetable}{lccc}
\tablecolumns{4}
\tablewidth{0pc}
\tablecaption{Best parameters for models of the Halo}
\tablehead{
\colhead{Model} & \colhead{$n$} & \colhead{$\rho_0$ (kpc$^{-3}$)} & \colhead{rms} }
\startdata
Spherical (c/a=1.0) & $-2.5\pm 0.1$ & $2.0\pm0.2$ & 0.14 \\
Constant flattening (c/a=0.6) & $-2.7\pm0.1$ & $6.6^{+0.8}_{-0.7}$ & 0.11 \\
Variable flattening & $-3.1 \pm 0.1$ & $4.2^{+0.5}_{-0.4}$ & 0.11 \\
\enddata
\label{tab-fit}
\end{deluxetable}

\clearpage
\begin{deluxetable}{cccccccccclcl}
\rotate
\tablecolumns{13}
\tablewidth{0pc}
\tablecaption{The Most Significant Overdensities}
\tablehead{
\colhead{Group} & \colhead{$n_{obs}$} & \colhead{$\alpha$} & \colhead{$\delta$} &
\colhead{$l$} & \colhead{$b$} & \colhead{$\langle V_0 \rangle$} & \colhead{$r_\sun$} &
\colhead{$R_{gc}$} & \colhead{$P_{ave}$} & \colhead{Confidence\tablenotemark{a}} &
\colhead{$F$\tablenotemark{b}} & \colhead{Comments} }
\startdata
1 & 6   & 72  & -1.0 & 199 & -28 & 15.6 & 10.2 & 17.5 & $4.6\times 10^{-6}$ & high   & 0.0090 & Mon Stream? \\
2 & 7   & 82  & -0.9 & 204 & -19 & 16.9 & 18.8 & 26.0 & $7.9\times 10^{-4}$ & low    & 1.1521 & \\  
3 & 12  & 128 & -1.3 & 226 & 22  & 15.7 & 10.5 & 16.8 & $2.7\times 10^{-5}$ & medium & 0.0101 & Mon Stream? \\
4 & 42  & 189 & -0.8 & 296 & 62  & 16.7 & 17.0 & 17.2 & $1.4\times 10^{-4}$ & low    & 0.0032 & Virgo feature \\
5 & 104	& 214 & -1.1 & 342 & 55  & 19.0 & 50.2 & 46.5 & $1.9\times 10^{-6}$ & high   & 0.0000 & Sgr Stream \\
6 & 6   & 230 & -0.5 & 2   & 45  & 17.3 & 22.3 & 17.6 & $7.3\times 10^{-4}$ & low    & 1.3848 & close to Pal 5 \\
\enddata
\tablenotetext{a}{The confidence of the group based on $P_{ave}$.}
\tablenotetext{b}{It provides an estimate of the frequency with which random variations 
alone produce groups of equal or greater significance than the observed group.}
\label{tab-overden}
\end{deluxetable}


\begin{thebibliography}{} 

\bibitem[Amrose \& McKay(2001)]{amr01} Amrose, S. \& McKay, T. 2001, \apjl,
560, L151

\bibitem[Bersier \& Wood(2002)]{ber02} Bersier, D. \& Wood, P. R. 2002, \aj, 123, 840

\bibitem[Brocato et al.(1996)]{bro96} Brocato, E., Castellani, V. 
\& Ripepi, V. 1996, \aj, 111, 809

\bibitem[Brown et al.(2005)]{bro05} Brown, W. R., Geller, M. J., Kenyon, S. J., 
Kurtz, M. J., Allende Prieto, C., Beers, T. C. \& Wilhelm, R. 2005, \aj, 130, 1097

\bibitem[Bullock et al.(2001)]{bul01} Bullock, J. S., 
Kravtsov, A. V. \& Weinberg, D. H. 2001, \apj, 548, 33

\bibitem[Bullock \& Johnston(2005)]{bul05} Bullock, J.S. \& 
Johnston, K. V. 2005, \apj, 635, 931

\bibitem[Cacciari(2003)]{cac03a} Cacciari, C. 2003, ASPC 296, 329

\bibitem[Cacciari \& Clementini(2003)]{cac03b} Cacciari, C. \& Clementini, G. 
2003, Lect. Notes Phys., 635, 105

\bibitem[Carraro \& Lia(2000)]{car00} Carraro, G. \& Lia, C. 2000, \aa, 357, 957

\bibitem[Carraro et al.(2005)]{car05} Carraro, G., V\'azquez, R. A., Moitinho, A. \&
Baume, G. 2005, \apj, 630, L153

\bibitem[Chaboyer(1999)]{cha99} Chaboyer, B. 1999, in Post-Hipparcos Cosmic Candles, 
Eds. A. Heck \& F. Caputo, Kluwer Ac. Pub., 111

\bibitem[Clement \& Shelton(1997)]{cle97} Clement, C. M. \& Shelton, I. 1997, \aj, 113, 1711

\bibitem[Conn et al.(2005a)]{con05a} Conn, B. C., Lewis, G. F., Irwin, M. J., 
Ibata, R. A., Ferguson, A. M. N., Tanvir, N. \& Irwin, J. M. 2005, \mnras, 362, 475

\bibitem[Conn et al.(2005b)]{con05b} Conn, B. C., Martin, N. F., Lewis, G. F., 
Ibata, R. A., Bellazzini, M. \& Irwin, M. J. 2005, \mnras, 364, L13

\bibitem[Cseresnjes(2001)]{cse01} Cseresnjes, P. 2001, \aap, 375, 909

\bibitem[Dall'Ora et al.(2003)]{dal03} Dall'Ora, M. et al. 2003, \aj, 126, 197

\bibitem[Dohm-Palmer et al.(2001)]{doh01} Dohm-Palmer, R. C. et al. 2001,
\apjl, 555, 37

\bibitem[Duffau et al.(2006)]{duf06} Duffau, S., Zinn, R., Vivas, A. K., 
Carraro, G., M\'endez, R. A., Winnick, R. \& Gallart, C. 2006, \apj, 636, L97

\bibitem[Freeman \& Bland-Hawthorn(2002)]{fre02} Freeman, K. \&
Bland-Hawthorn, J. 2002, \araa, 40, 487

\bibitem[Gnedin \& Ostriker(1997)]{gne97} Gnedin, O. Y. \& 
Ostriker, J. P. 1997, \apj, 474, 223

\bibitem[Grillmair et al.(1995)]{gri95} Grillmair, C. G., Freeman, K.C., 
Irwin, M. \& Quinn P.J. 1995, \aj, 109, 2553

\bibitem[Harding et al.(2001)]{har01} Harding, P., Morrison, H.L., 
Olszewski, E. W., Arabadjis, J., Mateo, M., Dohm-Palmer, R.C., 
Freeman, K.C. \& Norris, J.E. 2001, A.J., 122, 1397

\bibitem[Harris(1996)]{har96} Harris, W.E. 1996, \aj, 112, 1487

\bibitem[Hawkins(1984)]{haw84} Hawkins, M. R. S. 1984, \mnras, 206, 433 

\bibitem[Held et al.(2001)]{hel01} Held, E. V., Clementini, G., Rizzi, L., Momany, Y., Saviane, I. \& Di Fabrizio, L. 2001, \apj, 562, L39

\bibitem[Ibata et al.(2001)]{iba01}  Ibata, R., Lewis, G., Irwin,
M., Totten, E. \& Quinn, T. 2001a, \apj, 551, 294

\bibitem[Ibata et al.(2003)]{iba03}  Ibata R., Irwin, M., Lewis,
G., Ferguson, A. \& Tanvir, N. 2003, \mnras, 340, L21

\bibitem[Ideta \& Makino(2004)]{ide04} Ideta, M. \& Makino, J. 2004, \apj, 616, L107 

\bibitem[Irwin \& Hatzidimitriou(1995)]{irw95} Irwin, M. \& Hatzidimitriou, D.
1995, \mnras, 277, 1354

\bibitem[Ivans et al.(2000)]{iva00} Ivans, I. I., Sneden, C., Kraft, R. P., 
Suntzeff, N. B., Smith, V. V., Langer, G. E., \& Fulbright, J. P. 
2001, RevMexAA (Serie de Conferencias) 10, 21

\bibitem[Ivezi\'c et al.(2000)]{ive00} Ivezi\'c, \v{Z}. et al. 2000, \aj, 120, 9631

\bibitem[Ivezi\'c et al.(2004)]{ive04} Ivezi\'c, \v{Z}. et al. 2004, in 
ASP Conf Ser. 327, Satellite and Tidal Streams, ed. F. Prada,
D. Martinez-Delgado \& T. Mahoney, 104

\bibitem[Ivezi\'c et al.(2005)]{ive05} Ivezi\'c, \v{Z}., Vivas, A. K.,
Lupton, R. \& Zinn, R. H., 2005, \aj, 129, 1096

\bibitem[Johnston et al.(1996)]{joh96} Johnston, K. V., 
Hernquist, L., \& Bolte, M. 1996, \apj, 465, 278

\bibitem[Juric et al.(2005)]{jur05} Juric, M. 2005, astro-ph/0510520

\bibitem[Kaluzny et al.(1995)]{kal95} Kaluzny, J., Kubiak, M., Szymanski, M., 
Udalski, A., Krzeminski, W. \& Mateo, M. 1995, \aaps, 112, 407

\bibitem[Kinemuchi et al.(2002)]{kin02} Kinemuchi, K., Smith, H. A., Lacluyz\'{e}, A. P.,
Clark, C. L., Harris, H. C., Silbermann, N. \& Snyder, L. A. 2002, in Radial and Nonradial
Pulsations as Probes of Stellar Physics, eds. C. Aerts, T. R. Bedding and J. Christensen-Dalsgaard, ASP Conf. Series, 259, 130

\bibitem[King(1962)]{kin62} King, I. 1962, \aj, 67, 471

\bibitem[Kinman et al.(1965)]{kin65} Kinman, T. D., Wirtanen,
C. A. \& Janes, K. A. 1965, \apjs, 11, 223

\bibitem[Kinman et al.(1966)]{kin66} Kinman, T. D., Wirtanen,
C. A. \& Janes, K. A. 1965, \apjs, 13, 379

\bibitem[Kinman et al.(1982)]{kin82} Kinman, T. D., 
Mahaffey, C. T. \& Wirtanen, C. A. 1982, \aj, 87, 314
 
\bibitem[Kinman et al.(1985)]{kin85} Kinman, T. D., Kraft, R. P., 
Friel, E. \&  Suntzeff, N. B. 1985, \aj, 90, 95

\bibitem[Kinman et al.(1994)]{kin94} Kinman, T. D.,
Suntzeff, N. B. \& Kraft, R. P. 1994, \aj, 108, 1722

\bibitem[Kinman et al.(2004)]{kin04} Kinman, T. D., Saha, A. \& Pier, J. R.
2004, \apj, 605, L25

\bibitem[Kraft \& Ivans(2003)]{kra03} Kraft, R. P. \& Ivans, I. I. 2003,
\pasp, 115, 143

\bibitem[Kukarkin(1973)]{kuk73} Kukarkin, B. V. 1973, in Variable Stars in
Globular Clusters, IAU Colloquium 21, ed. J. D. Fernie, 8

\bibitem[Layden(1998)]{lay98} Layden, A. C. 1998, \aj, 115, 193

\bibitem[Lee et al.(2003)]{lee03} Lee, M. G. et al. 2003, \aj, 126, 2840

\bibitem[Leon et al.(2000)]{leo00} Leon, S., Meylan, G.
\& Combes, F. 2000, \aap, 359, 907

\bibitem[Liu \& Janes(1990)]{liu90} Liu, T. \& Janes, K. A. 1990, \apj, 360, 561

\bibitem[Majewski et al.(2003)]{maj03} Majewski, S. R., Skrutskie, M. F.,
Weinberg, M. D. \& Ostheimer, J. C. 2003, \apj, 599, 1082

\bibitem[Majewski et al.(2004)]{maj04} Majewski, S. R. et al. 2004, \aj, 128, 245

\bibitem[Mart{\'\i}nez-Delgado et al.(2001)]{m-d01} Mart{\'\i}nez-Delgado, D.,
Aparicio, A., G\'mez-Flechoso, M. A. \&  Carrera, R. 2001, \apjl, 549, L199

\bibitem[Mart{\'\i}nez-Delgado et al.(2005)]{m-d05} Mart{\'\i}nez-Delgado, D.,
Butler, D. J., Rix, H-W.,  Franco, Y. I., Pe\~narrubia, J., Alfaro, E. J. \& Dinescu, D. I.
2005, \apj, 633, 205

\bibitem[Martin et al.(2004)]{mar04} Martin, N. F., Ibata, R. A., Bellazzini, M., Irwin, M. J.,
Lewis, G. F. \& Dehnen, W. 2004, \mnras, 348, 12

\bibitem[Mateo(1998)]{mat98} Mateo, M. 1998, \araa, 36, 435

\bibitem[McConnachie \& Irwin(2006)]{mcc06} McConnachie, A. W. \& Irwin, M. J. 2006,
\mnras, 365, 1263

\bibitem[Momany et al.(2004)]{mom04} Momany, Y., Zaggia, S. R., Bonifacio, P., Piotto, G.,
De Angeli, F., Bedin, L. R. \& Carraro, G. 2004, \mnras, 421, L29

\bibitem[Nemec et al.(1988)]{nem88} Nemec, J. M., Wehlau, A. \& Mendes de Oliveira, C.
1988, \aj, 96, 528

\bibitem[Newberg et al.(2002)]{new02} Newberg, H. J. et al. 2002, \apj,
569, 245

\bibitem[Odenkirchen et al.(2001)]{ode01} Odenkirchen, M. et al. 2001,
\apj, 548, L165

\bibitem[Odenkirchen et al.(2002)]{ode02} Odenkirchen, M., Grebel, E. K.,
Dehnen, W., Rix, H.-W. \& Cudworth, K. M.  2002, \aj, 124, 1497

\bibitem[Odenkirchen et al.(2003)]{ode03} Odenkirchen, M. et al. 2003,
\aj, 126, 2385 

\bibitem[Pe\~narrubia et al.(2005)]{pen05} Pe\~narrubia, J. et al. 2005, 
\apj, 626, 128

\bibitem[Plaut(1966)]{pla66} Plaut, L. 1966, Bull. Astron. Inst.
Netherlands Suppl., 1, 105

\bibitem[Plaut(1968)]{pla68} Plaut, L. 1966, Bull. Astron. Inst.
Netherlands Suppl., 2, 293

\bibitem[Plaut(1971)]{pla71} Plaut, L. 1971, A\&AS, 4, 73

\bibitem[Preston et al.(1991)]{pre91} Preston, G. W., 
Schectman, S. A. \& Beers, T. C. 1991, \apj, 375, 121

\bibitem[Pritzl et al.(2002)]{pri02} Pritzl, B. J., Armandroff, T. E., Jacoby, G. H. \& 
Da Costa, G. S. 2002, \aj, 124, 1464

\bibitem[Pritzl et al.(2004)]{pri04} Pritzl, B. J., Armandroff, T. E., Jacoby, G. H. \& 
Da Costa, G. S. 2004, \aj, 127, 318

\bibitem[Pritzl et al.(2005)]{pri05} Pritzl, B. J., Armandroff, T. E., Jacoby, G. H. \& 
Da Costa, G. S. 2005, \aj, 129, 2232

\bibitem[Reid(1993)]{rei93} Reid, M. J. 1993, \araa, 31, 345

\bibitem[Rey et al.(2004)]{rey04} Rey, S.-C.,  Lee, Y.-W., Ree, C. H., Joo, J.-M., 
Sohn, Y.-J. \& Walker, A. R. 2004, \aj, 127, 958

\bibitem[Rocha-Pinto et al.(2003)]{roc03} Rocha-Pinto, H. J., 
Majewski, S. R., Skrutskie, M. F. \&  Crane, J. D. \apj, 594, L115

\bibitem[Saha(1985)]{sah85} Saha, A. 1985, \apj, 289, 310

\bibitem[Sandage(1990)]{san90} Sandage, A. 1990, \apj, 350, 603

\bibitem[Searle \& Zinn(1978)]{sz78} Searle, L. \& Zinn, R. 1978, \apj,
225, 357

\bibitem[Schlegel et al.(1998)]{sch98} Schlegel, D. J., 
Finkbeiner, D. P. \& Davis, M. 1998, \apj, 500, 525

\bibitem[Siegel \& Majewski(2000)]{sie00} Siegel, M. H. \& Majewski, S. R.
2000, \aj, 120, 284

\bibitem[Sirko et al.(2004)]{sir04} Sirko, E. et al., 2004, \aj, 127, 899

\bibitem[Smith(1995)]{smi95} Smith, H. A. 1995, RR Lyrae Stars (Cambridge
Astrophysics Series, 27)

\bibitem[Suntzeff et al.(1991)]{sun91} Suntzeff, N. B., Kinman, T. D. 
\& Kraft, R. P. 1991, \apj, 367, 528

\bibitem[Tsuchiya et al.(2003)]{tsu03} Tsuchiya, T., Dinescu, D. I. \& 
Korchagin, V. I. 2003, \apj, 589, L29

\bibitem[Vivas et al.(2001)]{viv01} Vivas, A. K. et al. 2001, \apjl, 554, L33

\bibitem[Vivas(2002)]{viv02} Vivas, A. K. 2002, PhD Thesis, Yale University

\bibitem[Vivas \& Zinn(2003)]{viv03} Vivas, A. K. \& Zinn, R. 2003, in Variability
with Wide Field Imagers, MSAIt, 74, 928

\bibitem[Vivas et al.(2004)]{viv04} Vivas, A. K. et al. 2004, \aj, 127, 1158

\bibitem[Vivas et al.(2005)]{viv05} Vivas, A. K., Zinn, R., \& Gallart, C. 
2005, \aj, 129, 189

\bibitem[Walker(1994)]{wal94} Walker, A. R. 1994, \aj, 108, 555

\bibitem[Walker \& Nemec(1996)]{wal96} Walker, A. R. \& Nemec, J. M. 1996, \aj, 112, 2026

\bibitem[Wesselink(1987)]{wes87} Wesselink, T. 1987, A Photometric Study
of Variable Stars in a Field near the Galactic Ceneter 
(Nijmegen:Brakkesnstein) 

\bibitem[Wetterer et al.(1996)]{wet96a} Wetterer, C. J., McGraw, J. T.,
 Hess, T. R. \& Grashuis, R. 1996, \aj, 112, 742

\bibitem[Wetterer \& McGraw(1996)]{wet96b} Wetterer, C. J. \& McGraw,
J. T. 1996, \aj, 112, 1046

\bibitem[Yanny et al.(2000)]{yan00} Yanny, B. et al. 2000, \apj, 540, 825

\bibitem[Yanny et al.(2003)]{yan03} Yanny, B. et al. 2003, \aj, 588, 824

\bibitem[Zinn \& West(1984)]{zin84} Zinn, R. \& West, M. J. 1984, \apjs,
55, 45

\bibitem[Zinn(1985)]{zin85} Zinn, R. 1985, \apj, 293, 424

\bibitem[Zinn(1986)]{zin86} Zinn, R. 1986, in Stellar Populations, 
Cambridge University Press, 73

\bibitem[Zinn et al.(2004)]{zin04} Zinn, R., Vivas, A. K., Gallart, C. \&
Winnick, R. 2004, in ASP Conf Ser. 327, Satellite and Tidal Streams, ed. F. Prada,
D. Martinez-Delgado \& T. Mahoney, 92

\end{thebibliography}
\end{document}